\documentclass[useAMS,usenatbib]{mn2e}
\usepackage{amsmath,amsfonts,amssymb,ulem,epsfig,times}
\usepackage[colour]{optional}
\bibpunct[, ]{(}{)}{;}{a}{}{,}

\title[Physical interpretation of the near-IR colours of low redshift galaxies]
{Physical interpretation of the near-infrared colours of low redshift galaxies}

\author[C.~Eminian et al.]
{C.~Eminian$^{1}$, G.~Kauffmann$^{2}$, S.~Charlot$^{3}$, V.~Wild$^{2}$,
 G.~Bruzual$^{4}$, 
 \newauthor A.~Rettura$^{5}$, J.~Loveday$^{1}$\\
$^{1}$ Astronomy Centre, University of Sussex, Brighton BN1 9QH, UK\\
$^{2}$ Max-Planck-Institut fuer Astrophysik,
Karl-Schwarzschild-Strasse 1, D-85748 Garching b. Muenchen, Germany\\
$^{3}$ Institut d'Astrophysique de Paris, CNRS, 98 bis Boulevard Arago, 
75014 Paris, France\\
$^{4}$ Centro de Investigaciones de Astronomia, AP 264, Merida 5101-A, Venezuela\\
$^{5}$ Department of Physics and Astronomy, Johns Hopkins University, 3400 N. Charles Street, Baltimore, MD 21218 USA\\
(C.Eminian@sussex.ac.uk)\\}

\begin{document}

\maketitle

\begin{abstract}

We use empirical techniques to 
interpret the near-infrared (near-IR) colours of a sample of 
5800 galaxies drawn from Sloan Digital Sky Survey (SDSS) 
main spectroscopic sample 
with $YJHK$ photometry from
the UK Infrared Deep Sky Survey (UKIDSS) data release one. Our study focuses on the inner 
3 arcsec regions of the galaxies sampled
by the SDSS fibre spectra. 
We study correlations between           
near-IR colours measured within this aperture 
and physical parameters derived from
the spectra. These parameters include specific star formation rate (SFR), 
stellar age,
metallicity and dust attenuation. 
All correlations are analyzed for samples of galaxies 
that are closely matched in
redshift, in stellar mass and in concentration index.       
Whereas more strongly star-forming galaxies have bluer optical colours,
the opposite is true at near-IR wavelengths -- galaxies with higher 
specific star formation rate have {\it redder} near-IR colours.
This result agrees qualitatively with
the predictions of models in which Thermally Pulsing Asymptotic Giant Branch
(TP-AGB) stars 
dominate the $H$ and $K$-band light of
a galaxy following a burst of star formation. 
We also find a surprisingly strong correlation between the near-IR colours 
of star-forming galaxies and their dust attenuation as   
measured from the Balmer decrement.
Unlike optical colours, however, near-IR colours 
exhibit very little dependence on galaxy inclination.
This suggests that the correlation of near-IR colours with dust attenuation
arises because TP-AGB stars are the main {\it source} of dust in the galaxy.  
Finally, we compare the near-IR           
colours of the galaxies in our sample to the predictions of three different
stellar population models: the \citet{Bruzual2003} model,
a preliminary version of a new model under development by
Charlot \& Bruzual, which includes a new prescription
for AGB star evolution, and the \citet{Maraston2005} model. 

\end{abstract}



\section{Introduction} \label{sect:intro}

Unlike optical colours, the near-IR colours of 
star-forming galaxies 
are poorly understood. The area and depth of sky
covered by near-IR surveys has been 
steadily increasing in recent years, so it is 
important to understand the nature of the stars that contribute to the 
light emitted in the wavelength range from 1 to 2 $\mu$m. This is necessary
if near-IR magnitudes and colours are to be 
used as a tool to infer physical information 
about integrated stellar populations. 
As we will see in this paper, the near-IR colours of local
spiral galaxies turn out to be very sensitive to stars
that contribute almost no flux at optical wavelengths 
and that are difficult to model theoretically.

One of the first observational studies of the integrated 
optical/near-IR colours of star-forming galaxies 
was carried out by \citet{Aaronson1978}, who showed that 
galaxies lie along a well-defined
morphological sequence in the $UVK$ colour plane 
in a similar manner to the well-known $UBV$
colour-morphology relation. \citet{Frogel1985} studied the 
$UBVJHK$ colours of 19
late-type spiral galaxy nuclei and showed that 
there was {\it no correlation} between
their $UBV$ and $JHK$ colours. This implied that
different stars contribute to the blue and the infrared light. 
It was proposed that the blue light was dominated 
by young stars and that the infrared light
was dominated by an old and metal-rich stellar population similar to that found
in elliptical galaxies.

Stellar population synthesis techniques can be used to 
{\it predict} which stars contribute
most strongly to the near-IR luminosity of a galaxy as a function 
of its age. The 
results show that the near-IR fluxes of galaxies are dominated by cool stars
on the red giant branch at ages 
greater than 1-2 Gyr and by stars located on the 
asymptotic giant branch (AGB) at younger ages. In the last phase of their evolution,
the thermally-pulsing (TP) phase, low metallicity AGB stars are extremely bright and dominate the
K-band light of a galaxy even though they are few in number. This evolutionary phase
is very difficult to model and can be a source of considerable discrepancy between
different population synthesis models (\citealt{Maraston2006}, \citealt{Bruzual2007}). 

It has been proposed that the combination of optical and near-IR photometry 
can break the well-known degeneracy between the effects of age and metallicity
on the stellar populations of galaxies at optical wavelengths.
The idea is that the optical colours are primarily sensitive to age, 
because the main contribution to the optical luminosity of a galaxy 
comes from upper main sequence and 
turn-off stars. On the other hand the near-IR colours ($H$ and $K$ bands) are 
mainly determined by stars in the AGB phase at low metallicity and in the red giant branch stars (RGB)
phase at high metallicity.  
Low metallicity carbon-AGB stars are one 
magnitude brighter in the $K$-band than oxygen-rich M stars 
and the ratio 
of carbon-rich to oxygen-rich stars decreases sharply with increasing 
metallicity \citep{Lee2007}. Several studies have used the combination of near-IR and 
optical broad band colours to estimate the ages and metallicities of star-forming galaxies.
Generally a maximum likelihood method is used to infer these parameters from the 
nearly orthogonal age-metallicity 
grids that are predicted using stellar population synthesis models.

\citet{deJong1996} analysed radial gradients in the B-V and  
$r$-K colours of 86 face-on spiral galaxies and tested different hypotheses to 
explain their origin. His conclusion was that the 
colour gradients are best explained by 
a combined stellar age and metallicity gradient accross the disk, with the 
outer regions being younger and less metal-rich on average. Similar 
analyses have been conducted by \citet{Bell2000} and \citet{MacArthur2004} 
using samples of low inclination galaxies 
that span a range in Hubble type. They 
used $r$-H/K versus B-$r$ colours to investigate age and 
metallicity gradients and showed that both 
age and metallicity are strongly correlated 
with local surface brightness. Studying elliptical and spiral galaxies, 
\citet{Mobasher1986} have highlighted that near-IR colours are particularly 
sensitive to metallicity.
The most recent study has been performed by 
\citet{Wu2005} using a sample of 36 nearby early-type galaxies from the Sloan
Digital Sky Survey. This study found that metallicity 
variations are responsible for the
observed colour gradients in these systems.

These results have recently been challenged by \citet{Lee2007}, who show that the orthogonal 
age-metallicity grids computed using the 
\citet{Bruzual2003} models become much more degenerate when more        
accurate treatments of TP-AGB stars and of convective core overshoot are included.
However, even after these improvements, the models have 
difficulty in matching data on star clusters of known age and metallicity.
\citet{Geller2006} also encountered difficulties when attempting to model
the near-IR colours of interacting galaxies in close pairs. 
They found a subset of galaxies with much redder $H-K$ colours than could be
explained by simple models and interpreted their results as evidence for 
extremely hot (600-1000 K) dust within compact star-forming regions.
 
In this paper, we adopt an empirical approach.
We combine broadband $YJHK$ photometry from the UK Infrared Deep Sky 
Survey (UKIDSS) data release one with optical photometry from 
Sloan Digital Sky Survey (SDSS) fourth data release. We calculate optical
and near-IR colours within the 3 arcsecond diameter 
fibre aperture sampled by the SDSS spectra.
This allows us to study correlations between both
optical and near-IR colours and a variety of physical parameters
that have been derived directly from the spectra; 
these include specific star formation rate, metallicity,
dust attenuation, and mean stellar age. Because 
UKIDSS and SDSS spectroscopy are available for a sample of more than 5000
galaxies, we are able to study these correlations for subsamples
of galaxies that are closely matched in stellar mass and in
redshift.
We also compare the optical and near-IR colours of the galaxies
in our sample to the predictions of the 
stellar synthesis population models of \citet{Bruzual2003},
a preliminary version of an improved model under development by 
\citet{Charlot2007} and the models of \citet{Maraston2005}, 
which implement the
TP-AGB phase using "fuel consumption" approch.\\

This paper is organised as follows. We describe our optical and near-IR
data in section \ref{sect:data}, as well as the physical parameters used for our
analysis. Our methods for analyzing the correlations between 
near-IR colours and physical parameters 
 are presented in section 
\ref{sect:methods} and \ref{sect:results_cor}. In section \ref{sect:models}, 
we compare the near-IR colours of the galaxies in 
our sample with the colours predicted by 
stellar population models. Finally we discuss the implications of our results 
in section \ref{sect:discussion} and we summarize our findings in section
\ref{sect:summary}.


\section{The data} \label{sect:data}

\subsection{Optical photometry} \label{sect:optical}
The optical photometry is drawn from the Sloan Digital Sky 
Survey (SDSS) photometric galaxy catalogue \citep{York2000}. We restrict our sample to galaxies 
in the publically available MPA SDSS database 
(http://www.mpa-garching.mpg.de/SDSS), 
for which physical parameters have been derived from the galaxy spectra. 
Our optical sample contains       
473,034 galaxies from the main spectroscopic sample of the SDSS fourth data release 
(DR4; \citealt{Adelman_McCarthy2006}), covering 4783 square degrees to a depth 
of 17.77 in $r$-band Petrosian AB magnitude. The magnitudes of interest are 
measured through the fibre aperture 
in the $u$, $g$, $r$, $i$, $z$ filters. 
We have applied corrections for foreground Galactic
extinction according to \citet{Schlegel1998} and we have also {\it k}-corrected
the magnitudes to redshift 0.1 (the 
mean redshift of the sample) using v3 of the KCORRECT code \citep{Blanton2003} and assuming 
a standard cosmology with $\Omega=0.3$, $\Lambda=0.7$ and
$H_0 = 70$ km s$^{-1}$ Mpc$^{-1}$. Our sample covers a redshift range from 0.005 to 0.3. 
In the following sections, 
we will use the concentration parameter $C$, defined as the ratio of the radii 
enclosing 90 and 50 per cent of the galaxy light in the $r$-band, as an 
indicator of the morphological type of the galaxy. The axis ratio of the galaxy 
is defined as the ratio of the isophotal minor axis to the isophotal major axis in the 
$r$-band, where the isophotes are measured at 25 magnitudes per square
arcsecond. 


\subsection{Near infrared photometry} \label{sect:nearIR}
\begin{figure}
\begin{center}
\includegraphics[width=0.48\textwidth]{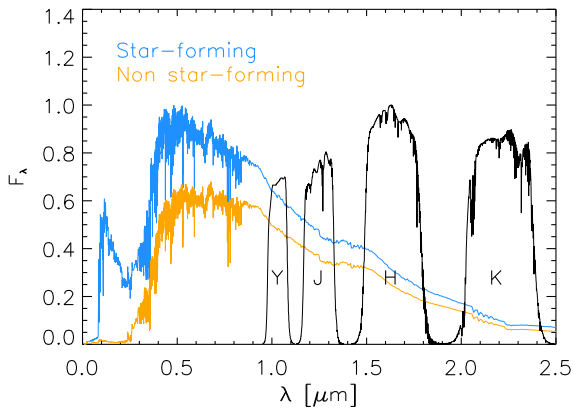}
\caption{Response functions of the
UKIDSS $YJHK$ filters (black curves) are plotted on top of spectra
computed for a star-forming galaxy (blue) and for a non star-forming 
galaxy (orange) at $z$ = 0.1 using the \citet{Bruzual2003} models.} 
\label{fig:fig1}
\end{center}
\end{figure}

The near-IR data come from the UK Infrared Deep Sky Survey 
(\citealt{Lawrence2007}). The Large Area Survey, the largest of the five surveys 
that 
constitute UKIDSS, is the near-IR counterpart of SDSS. The images are 
obtained by the Wide-Field Camera on the UK Infrared Telescope in 
Hawaii. At the end of the 7 year observing campaign, it will cover 4000 square degrees of 
the Northern sky in the $YJHK$ filters and it will reach a limiting $K$-band
magnitude that is three magnitudes deeper than that of 2MASS
\citep{Skrutskie2006}. 
In the present study we focus on galaxies contained in the
first data release of the Large Area Survey \citep{Warren2007}. This
reaches a depth of 18 in $K$-band Vega magnitude and covers 320 square
degrees of the Northern sky. Our near-IR sample comprises 
2.5 million objects with measured $K$-band fluxes. The four filters 
cover the near infrared wavelenths from 1 micron for the $Y$ filter to 2.4 
micron for the $K$ filter. Circular 
aperture photometry is available for 13 diameters. We extract the 
Vega magnitude measured through a 2.8 arcsec diameter aperture \footnote{From 
the UKIDSS catalogue apermag4 magnitudes were selected, which have 
been corrected for the effects of aperture and seeing under the 
assumption that the object is a point source. However, we have verified 
that the resulting colours are unaffected by this correction}, 
which is well-matched to the 3 arcsec diameter fibre aperture of SDSS. 
These magnitudes are corrected for Galactic extinction 
and {\it k}-corrected to redshift 0.1. In Fig. \ref{fig:fig1},
we show the response
function of the UKIDSS $YJHK$ filters on top of spectra computed for a
10 Gyr old star-forming galaxy (blue) and for a 10 Gyr old non star-forming 
galaxy (orange) at $z$ = 0.1 using the Bruzual \& Charlot (2003) models.

\subsection{Physical parameters} \label{sect:physics}

In this section, we briefly define the physical parameters that are
used in our analysis: specific star formation rate,
age, stellar and interstellar metallicity and dust attenuation. These quantities 
are all available in the public SDSS MPA database\footnote{
http://www.mpa-garching.mpg.de/SDSS/} and have been derived from 
the galaxy spectra. We note that the fibres in the SDSS spectrograph
have diameters of 3 arcsec.
At the mean redshift of our sample
this corresponds to a diameter of 7 kpc (h = 0.7). Analyzing galaxy properties within
a fixed fibre aperture means that the physical size of the region for which
we carry out our analysis will be larger at higher redshifts.

\begin{table*}
\begin{center}
\caption{Values of the physical parameters for the star-forming and non
star-forming samples}  
\label{tab:phys_sf}
\begin{tabular}{lccccccc}
\hline
\hline
 &M$_{{\rm fib}}$ & SFR/M$^{*}$ & Age & Z$_{{\rm gas}}$ & Z$_{{\rm stellar}}$ & H$\alpha$/H$\beta$
 &A$_{z}$\\
 & [log M$_{\odot}$] & [log yr$^{-1}$] & [Gyr] & [12+log(O/H)] & [$Z_{\odot}$] &
 - & [mag] \\
\hline
SF: Mean value & 9.6 & -9.9 & 2 & 9.0 & 0.7 & 4.1 & 0.6 \\
SF: 1 sigma  & 0.7 & 0.4 & 1 & 0.2 & 0.5 & 0.7 & 0.3 \\
\hline
NSF: Mean value & 10.1 & - & 7 & - & 1.1 & - & 0.0 \\
NSF: 1 sigma  & 0.5 & - & 2 & - & 0.4 & - & 0.2 \\
\hline
\hline
\end{tabular}
\end{center}
\end{table*}

\subsubsection{Star-forming and non star-forming classes} \label{sect:sf_nsf}

In order to select a sample of star-forming
galaxies free of AGN contamination, 
we use the emission-line 
classifications of \citet{Brinchmann2004}. The classifications are based on the
location of galaxies in the Baldwin, Phillips \& 
Terlevich diagram \citep{Baldwin1981}. Galaxies are divided into 6 classes: star-forming, composite,
AGN, low S/N AGN, low S/N star-forming, and unclassifiable. We selected the 
star-forming and the non star-forming class (unclassifiable) for our analysis.
The average values of the physical parameters are listed in 
Table\ref{tab:phys_sf} for these two classes. 
As noted by \citet{Brinchmann2004}, it is important to remember that 
this is a nuclear classification for the galaxies lying at redshift 0.1 and
below, as these measurements are restricted to the central regions of
the galaxies.

\subsubsection{Star formation rate} \label{sect:sfr}

The star formation rate (SFR) measured inside the fibre is also taken from the
\citet{Brinchmann2004} analysis. To estimate this parameter, these authors
matched the intensity of 
a variety of emission lines, including H$\alpha$ and H$\beta$,
to a grid of HII region photo-ionization models.
Because the stars that
contribute to the ionizing flux are massive and very young, with a lifetime $<$
20 Myr, this method provides a measure of the instantaneous star formation 
rate, independent of the previous star formation history. 
For our study, we use the specific SFR
measured inside the fibre (SFR/M$^{*}$).

\subsubsection{Age and metallicity}\label{sect:age_met}

We use the light-weighted stellar ages and the stellar metallicities
estimated by \citet{Gallazzi2005} using a 
carefully chosen set of absorption features that can break the age-metallicity
degeneracy and that depend weakly on the $\alpha$/Fe abundance ratio: D4000, H$\beta$, 
H$\delta_{A}$+H$\gamma_{A}$, [Mg$_{2}$Fe] and [MgFe]'. While the composite Mg+Fe
indices are a good indicator of metallicity, the H-Balmer lines are mostly
sensitive to age. For star-forming galaxies, we use 
the gas-phase oxygen abundance derived by \citet{Tremonti2004} 
from the strong optical nebular emission lines as the metallicity indicator, 
because for these galaxies, 
the gas-phase metallicity can be measured with better
precision than the stellar metallicity.
These values are expressed in 12 + log (O/H), where the solar metallicity in 
these units is 8.69. 

\subsubsection{Dust} \label{sect:dust}

As an estimate of the dust content of our star-forming galaxies, we 
have used the ratio of the H$\alpha$ to the H$\beta$ emission line fluxes. 
These emission lines are measured after the 
subtraction of the continuum spectra. As H$\alpha$ and H$\beta$ are attenuated
differently because of their different wavelengths, their ratio is a 
measure of the amount of dust present in and around HII regions, where the star
formation occurs. We assume a value of 2.86
corresponds to the dust-free case. An independent 
measurement of the dust content is provided by the
$z$-band attenuation parameter $A_z$  
derived by \citet{Kauffmann2003}, which measures the amount of dust in 
the interstellar medium of the galaxy. It is calculated by fitting spectra
with stellar population synthesis models with an attenuation curve of the
form $\tau_{\lambda}$ $\propto$ $\lambda^{-0.7}$. The parameter $A_z$ can be 
estimated for both star-forming and non star-forming galaxies.


\subsection{Matching near-IR and optical} \label{sect:match_data}

We have matched the SDSS and UKIDSS samples in right ascension 
and declination with a maximal allowed 
separation of 0.5 arcsec in both coordinates. This upper limit is consistent with the 
measurement errors on the position of SDSS and UKIDSS, where 96\% of SDSS objects
have an error $<$ 0.2 arcsec in right ascension and declination and all UKIDSS objects
have an error $<$ 0.26 arcsec. The mean separation obtained for the match is 0.17
arcsec. The matched sample consists of 12,415 galaxies with detections in all
four near-IR bands. Nearly all SDSS galaxies with spectroscopy 
that are in the field observed by UKIDSS have 
a near-IR counterpart. The  
star-forming and non star-forming galaxy classes consist of
2377 and 3438 galaxies respectively.


\section{Methods} \label{sect:methods}

In this section we present our method to derive the correlation
between the optical/near-IR colours and the physical 
parameters described in section \ref{sect:physics}. 
We analyse the star-forming and the non star-forming 
galaxies separately. The pair matching method, described in the following
section, allows us to suppress the dominant correlation with stellar mass.
For each class, we then evaluate the 
correlation between the optical/near-IR colours and specific SFR, 
stellar age, metallicity and dust attenuation of paired galaxies using the Pearson 
correlation coeficient. 

\subsection{Matched galaxy pairs} \label{sect:pair_matching}

It is well known that colours are strongly 
correlated with galaxy mass; the most massive galaxies have redder optical
colours and the same is true for
near-IR colours. The main physical driver of this
trend is the fact that low-mass galaxies are currently forming
stars and high-mass galaxies have very little ongoing star formation
\citep{Kauffmann2003b}. If we wish to isolate the effects of other 
physical parameters on colours, it is useful to remove this             
dominant trend by comparing galaxies with a fixed stellar mass.
We use a pair-matching technique; for each galaxy, we find a "partner"
that is closely matched in stellar mass, in redshift
and in concentration index. 
We require a close match in redshift, because for a given stellar mass, the sample
is biased towards bluer galaxies at higher redshifts. Matching in redshift also
ensures that we compare colours within the same physical radius.  
Matching in concentration means that we 
compare galaxies with the same morphology, because the colour is
influenced by the presence of a bulge component in the galaxy.
To summarize, our matched galaxy pairs have: 
$\Delta$log$M_{{\rm fib}}$ $<$ 0.05 with $M_{{\rm fib}}$ 
expressed in M$_{\odot}$, $\Delta$z $<$ 0.02 and $\Delta$C $<$ 0.1.

We proceed
as follows: we begin by randomising the order of the galaxies in our sample. 
Then we take the first galaxy and define its partner as the galaxy minimising the
quantity $sep$ defined by equation~(\ref{equ:sep}) below and satisfying the limits in 
$\Delta$log$M_{{\rm fib}}$, $\Delta$z and $\Delta$C defined above, where the
denominators are the average of the absolute values. After this
operation, the two galaxies are removed from the list, and we repeat the
procedure with the rest of the sample.
\begin{equation}\label{equ:sep}
{\rm sep} = \frac{|\Delta log M_{{\rm fib}}|}{{\rm avg}(|\Delta log M_{{\rm fib}}|)} + 
\frac{|\Delta z|}{{\rm avg}(|\Delta z|)} +
\frac{|\Delta C|}{{\rm avg}(|\Delta C|)} 
\end{equation}
This results in 708 and 673 galaxy pairs for the 
star-forming and non star-forming classes, respectively. The pairs
are required to have unproblematic 
measurements of mass, specific SFR (for the star-forming class only), 
age, metallicity and dust attenuation for both galaxies.


\subsection{Calculation of the correlations} \label{sect:correlation} 

We use our pair sample to study correlations between 
the colour differences of 
paired galaxies and differences in 
specific SFR, stellar age, metallicity and dust attenuation. 
The degree of correlation is 
evaluated using the Pearson's product-moment 
correlation coefficient $r$, defined as
\begin{equation}\label{equ:cor}
r (x,y) = \frac{{\rm cov} (x,y)}{\sigma (x) \times \sigma (y)} \,.
\end{equation}
where $\sigma$ is the standard deviation. The correlation coefficient measures the degree of linear dependence 
between the two variables. When they are independent, then this coefficient is 
zero. In general, $r$ varies between $-$1 and +1, where $r$ = $-$1 means that $x$ 
and $y$ are
linearly dependent with a negative slope, and $r$ = +1 means that they are 
linearly dependent with a positive slope. 
The errors on the correlation coefficients are calculated analytically by combining the
measurement errors of the colours and the errors of the
derived physical parameters (see Brinchmann et al 2004 for more details).
We also use a resampling technique to assess whether the derived value
of $r$ depends on the choice of galaxy pairs.
The final values of $r$ and their associated errors 
represent the average of 1000 resamplings of the pair-matching process. 


\section{Results for the correlation} \label{sect:results_cor}

\subsection{Correlations for star-forming galaxies}\label{sect:cor_sf}

\begin{figure}
\begin{center}
\includegraphics[width=0.5\textwidth]{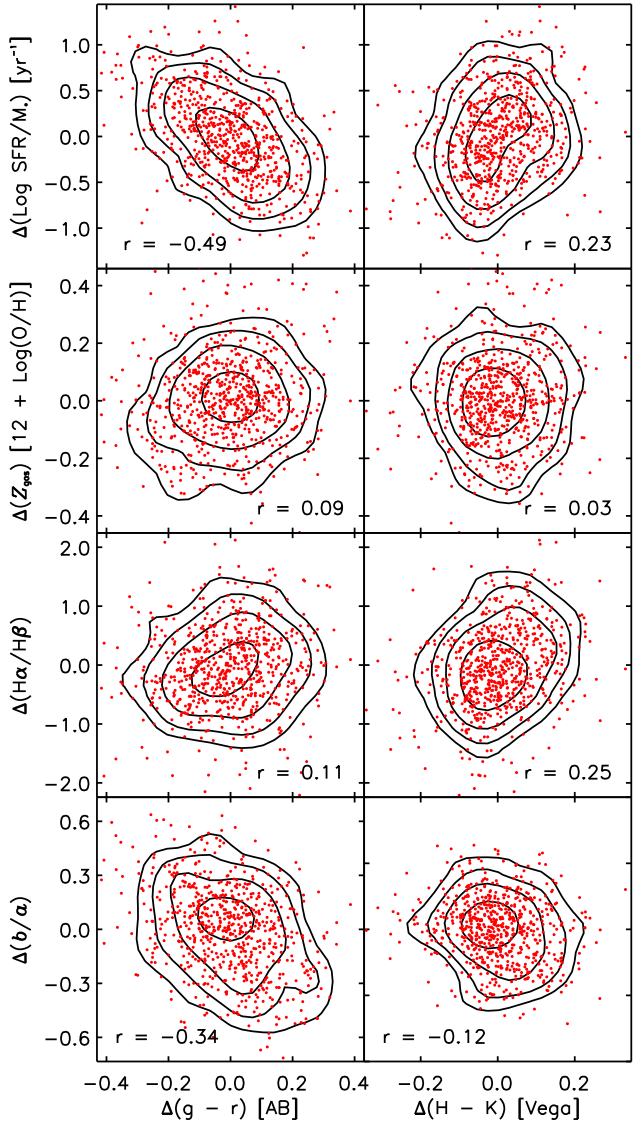}
\caption{Correlation between the differences in the specific SFR,
metallicities, Balmer decrements and axis ratios of the galaxy pair and the differences in their 
$g-r$ and $H-K$ colours. The contours indicate the density of points. 
Results are shown for the star-forming 
sample.}  
\label{fig:fig2}
\end{center}
\end{figure}

The results of our analysis for the star-forming sample are summarised in 
Table\ref{tab:cor_sf}. Given the null hypothesis that they are uncorrelated and taking into account
the size of the samples, the probability that $|r|$ should be larger than 0.074
is ~5\% i.e. correlation coefficients below this value should be regarded as
insignificant. The correlation coefficients have been calculated for
optical and near-IR colours as a function of 
specific SFR, mean stellar age, gas-phase
metallicity, H$\alpha$/H$\beta$ ratio, $z$-band attenuation, and
isophotal axis ratio. We note that all these parameters 
except the last one are
derived from the galaxy spectra. 
Results are listed for the optical colours
$g-r$ and $i - z$ and the infrared colours 
$Y-J$ and $H-K$. The chosen set of colours 
enables us to study how the shape of a galaxy's        
spectral energy distribution depends on
physical parameters from the optical to the near-IR 
in a continuous way. 
We also show examples of these correlations in 
Fig.\ref{fig:fig2}:
$g-r$ and $H-K$ colour differences as a function of differences
in specific SFR, gas-phase metallicity, H$\alpha$/H$\beta$ ratio
and isophotal axis ratio.

\begin{table*}
\begin{center}
\caption{Correlation coefficients for star-forming galaxies for the relations
between optical/near-IR colours and the following physical parameters:
SFR/M$^{*}$ (the specific SFR measured inside the
fibre), mean stellar age, Z$_{{\rm gas}}$ (the gas-phase metallicity), the
H$\alpha$/H$\beta$ ratio (Balmer decrement),
$A_z$ (the attenuation of the $z$-band light by dust), and $b$/$a$ (the
isophotal axis ratio). The errors are estimated as described in the text.}  
\label{tab:cor_sf}
\begin{tabular}{ccccccc}
\hline
\hline
\multicolumn{7}{c}{Correlation Coefficient}\\ 
\hline
Colour & SFR/M$^{*}$ & Age & $Z_{{\rm gas}}$ & H$\alpha$/H$\beta$ & $A_{z}$ & $b$/$a$\\
\hline
$g-r$ & $-$0.49 $\pm$ 0.05 & 0.62 $\pm$ 0.09 & 0.09 $\pm$ 0.05 & 0.11 $\pm$ 0.03 & 
$-$0.09 $\pm$ 0.03 & $-$0.34 $\pm$ 0.04 \\ 
$i - z$ & $-$0.37 $\pm$ 0.05 & 0.46 $\pm$ 0.08 & 0.06 $\pm$ 0.05 & 0.13 $\pm$ 0.03 & 
$-$0.04 $\pm$ 0.04 & $-$0.31 $\pm$ 0.04 \\ 
$Y-J$  & $-$0.02 $\pm$ 0.02 & 0.07 $\pm$ 0.05 & 0.08 $\pm$ 0.03 & 0.20 $\pm$ 0.03 & 
0.12 $\pm$ 0.03 & $-$0.20 $\pm$ 0.02 \\ 
$H-K$  & 0.23 $\pm$ 0.03 & $-$0.17 $\pm$ 0.05 & 0.03 $\pm$ 0.03 & 0.25 $\pm$ 0.03 & 
0.23 $\pm$ 0.04 & $-$0.12 $\pm$ 0.03 \\
\hline
\hline
\end{tabular}
\end{center}
\end{table*}

\begin{table*}
\caption{Correlation coefficients for non-star-forming galaxies for the
relations between
optical/near-IR colours and the following physical parameters: 
mean stellar age, stellar metallicity, $A_z$ (the 
attenuation of the $z$-band light by dust), and $b$/$a$ (the isophotal axis
ratio).} 
\label{tab:cor_nsf}
\begin{tabular}{ccccc}
\hline
\hline
\multicolumn{5}{c}{Correlation Coefficient}\\
\hline
Colour & Age & $Z_{{\rm stellar}}$ & $A_{z}$ & $b$/$a$\\
\hline
$g-r$ & 0.4 $\pm$ 0.1 & 0.05 $\pm$ 0.08 & $-$0.02 $\pm$ 0.05 & $-$0.08 $\pm$ 0.11 \\ 
$i - z$ & 0.20 $\pm$ 0.09 & 0.00 $\pm$ 0.07 & 0.16 $\pm$ 0.05 & $-$0.08 $\pm$ 0.05 \\ 
$Y-J$ & 0.08 $\pm$ 0.04 & 0.04 $\pm$ 0.04 & 0.02 $\pm$ 0.02 & $-$0.04 $\pm$ 0.02 \\ 
$H-K$ & $-$0.04 $\pm$ 0.04 & 0.07 $\pm$ 0.04 & 0.14 $\pm$ 0.03 & $-$0.01 $\pm$ 0.03 \\
\hline
\hline
\end{tabular}
\end{table*}

As can be seen from Table\ref{tab:cor_sf} and from
Figure\ref{fig:fig2},
the $g-r$ colour is strongly anti-correlated with the specific SFR ($r$ = $-$0.49), 
with the bluest colours corresponding to the
most strongly star-forming galaxies. This
result is expected, because the emission from young stars peaks at UV and blue
wavelengths. This anti-correlation still exists for the
$i - z$ colour, but 
the $H-K$ colour displays a positive correlation with specific SFR ($r$ =
0.23 as seen in Fig.\ref{fig:fig2}). At a constant mass, 
strongly star-forming galaxies have {\it redder} 
near-IR colours. Even if this correlation is not as
strong as the one in the optical, it is significant. One possibility
is that that AGB stars are responsible 
for this correlation, because they dominate the $K$-band luminosity
of a galaxy 0.1 Gyr
after a burst of star formation (see section \ref{sect:bc03_cb07} for 
detailed predictions from stellar population models). Galaxies with
ongoing star formation are rich in AGB stars and therefore 
have redder near-IR
colours than quiescient galaxies of the same mass.

The $g-r$ colour also correlates with the mean stellar age measured from
stellar-absorption lines; galaxies 
with older ages have redder optical colours. 
This correlation becomes weaker at longer
wavelengths, and once again the sign of the correlation
reverses for the $H-K$ colour ($r$ = $-$0.17).
The gas-phase metallicity  
does not seem to correlate with any colour.
At a first glance, this is a surprising result, because
metal-rich stellar populations are predicted to have 
redder colours. 
The most natural explanation is that the metallicity of a galaxy 
is very strongly correlated with its mass and not with any other parameter
\citep{Tremonti2004}. Because our galaxy pairs are
constrained to have the same stellar mass, the correlation 
with metallicity is strongly suppressed. 

The correlation coefficients for the H$\alpha$/H$\beta$ ratio
and the $z$-band attenuation are also small 
for optical colours. Curiously, the strength of the 
correlation {\it increases} towards near-IR wavelengths ($r$ = 0.25 and
0.23 for the correlation between $H-K$ colour and H$\alpha$/H$\beta$ 
and $A_z$ respectively). This result is very surprising, as dust should  
affect optical colours more than near-IR colours.

One way to gain further insight into this result is to look
at the correlation between optical and near-IR colours and
the axis ratios of the galaxies (see Table\ref{tab:cor_sf}
and the bottom panels in Fig.\ref{fig:fig2}). Our results show that 
low-inclination ($b$/$a$ $\simeq$ 1) galaxies have bluer optical colours
than high-inclination ($b$/$a$ $\ll$ 1) galaxies. 
This is most likely due to dust reddening, because
the light coming from the center of high-inclination galaxies passes through
a large amount of material, whereas for low-inclination
galaxies it comes directly to us. 
As seen from Table\ref{tab:cor_sf} the correlation of the near-IR
colours with the axis ratios is much weaker.  
This leads us to suggest that 
the positive correlation of the near-IR colours with dust 
attenuation is caused by some phenomenon other than dust reddening. 
AGB stars are known to release large amounts of dust.
We thus hypothesize that 
galaxies with redder $H-K$ colours contain a higher fraction of AGB stars,
and {\it by extension} a larger amount of dust.
The apparent tendency for the near-IR colours to correlate with
Balmer decrement and the $z$-band attenuation is thus an induced
correlation, not a primary one.


\subsection{Correlations for non star-forming galaxies}\label{sect:cor_nsf}

Correlations calculated for the non star-forming galaxy sample 
are presented in 
Table\ref{tab:cor_nsf}. We show the correlation coefficient and its error  
for the relations between optical/near-IR colours and 
mean stellar age, stellar
metallicity, $z$-band dust attenuation, 
and the isophotal axis ratio. 
All the correlation coefficients are smaller than those
obtained for the star-forming sample, 
but the overall results are quite similar.
The strongest correlation is that between age and $g-r$ colour
($r$ = 0.4), and the strength of the correlation decreases at longer
wavelengths ($r$ = $-$0.04 for the $H-K$ colour). 
Table\ref{tab:cor_nsf} shows
that the correlation with stellar metallicity is below the 5\%
significance level ($r$ $\le$ 0.07). 
Once again, the correlation of colours with $z$-band dust 
attenuation increases at near-IR wavelenghs, while
the absolute correlation with isophotal axis ratio decreases. We note that
the correlation between $g-r$ colour and isophotal axis ratio is 
less significant for this sample, as most of the galaxies are 
ellipticals. 
 

\section{Comparison with stellar population models}\label{sect:models}

\subsection{Description of \citet{Bruzual2003} and \citet{Charlot2007} models}

We use the \citet{Bruzual2003} (hereafter BC03) stellar population synthesis
models to compute the optical and near-IR colours of model "galaxies" 
for comparison with the data. We also test here a preliminary version of
the \citet{Charlot2007} (hereafter CB07) code, which includes a new prescription
for AGB star evolution. We adopt the Padova 1994 single stellar population
evolutionary tracks and the Chabrier initial mass function. 
\footnote {The BC03 model package allows one to
choose between Padova 1994 and Padova 2000 evolutionary tracks, but the latter
tends to produce worse agreement with observed galaxy colours.
On the other hand,
as shown by \citet{Bruzual2003}, the choice of the 
initial mass function should not influence significantly the output colours.} 
The models allow us to compute the evolution of the spectrum 
and the colours of a composite stellar population between 0 and 20 Gyr 
for a given metallicity and star formation history. For each model galaxy, 
the output colours are evolution and {\it k}-corrected to $z$ = 0.1, 
so that they can be compared directly with 
the observational data. 

We have computed a grid of models with different
star formation histories and metallicities that are designed to
span the observed colours
of the galaxies in our sample. We parametrize an exponential declining
 SFR as
\begin{equation}\label{equ:sfr_exp}
\phi_{{\rm exp}}(t) = \frac{e^{-t / \tau}}{\tau},
\end{equation}
where $t$ is the age and $\tau$ is the star formation timescale.
To represent the star-forming sample, we have chosen two models.
The first one has a constant SFR, and the second one has an exponential
declining SFR with  
$\tau$ = 3 Gyr. In general, the star formation history of early-type 
spiral galaxies can be described by an exponential law, because the gas
available for forming new stars 
decreases with time \citep{Kennicutt1998}. 
The constant SFR model describes galaxies where the available gas is   
replenished through further infall, and 
is therefore suited to describe later-type galaxies with 
higher rates of ongoing star formation. To model our non star-forming
sample, we choose $\tau$ = 1 Gyr. In this model, the star formation 
rate decreases rapidly and reaches values close to zero at $t$ = 5 Gyr. 
This model is designed to reproduce the star formation histories of
non-interacting elliptical galaxies. 

We generate these models for five stellar metallicities: 0.02$Z_{\odot}$, 
0.2$Z_{\odot}$, 0.4$Z_{\odot}$, $Z_{\odot}$, 2$Z_{\odot}$. This range is
broader than that spanned by the galaxies in our samples, which
vary between 
0.2$Z_{\odot}$ and $\sim$$Z_{\odot}$. 
Note that our model galaxies have a fixed metallicity, i.e. 
there is no metal enrichment with age. 
We plot galaxy colours at model ages of 5, 10 and 15 Gyr (we expect the
typical age of the galaxies in our sample to be around 10 Gyr). 
The upper limit of 15 Gyr is older than currently favored estimates of the
age of the Universe, but
in a few cases models with large ages are needed to fit the data. We note
that by "age", we mean here the time for which the model galaxy has been
forming its stars; this must be distinguished from
the observed age of the stellar population (such as that considered in
section\ref{sect:age_met}),
which is usually a measure of its luminosity-weighted mean stellar age.
The luminosity-weighted age is generally
considerably lower, because 
young stars always dominate
the optical luminosity in a composite stellar population. 

The effect of dust attenuation on the colours is included using the simple 
two-component model of \citet{Charlot2000}. This accounts for different 
attenuations affecting young stars in their birth clouds (i.e., giant
molecular clouds) and older stars in the diffuse ambient (i.e. diffuse)
interstellar medium of a galaxy. The main two adjustable parameters of this model
are the total effective V-band optical depth affecting stars younger than 10$^{7}$ yr
(the typical lifetime of a giant molecular cloud), $\hat{\tau}_{V}$, and
the fraction of that optical depth that is contributed by dust in the ambient 
interstellar medium, $\mu$.
Our models include dust attenuation 
with $\hat{\tau}_{V}$ = 1 and 
$\mu$ = 0.3. This corresponds to an attenuation of 1.1 magnitude in the V-band 
for stars younger than 10$^{7}$ yr and 0.3 mag for 
older stars. We note that the estimated 
$z$-band attenuation varies between 0.4 magnitude and 0.8 
magnitude for the star-forming galaxies in our sample and between 0. and 0.03 for 
the non star-forming galaxies. The attenuation curve across the spectrum has
the form $\tau_{\lambda}$ = $\tau_{V}$ ($\lambda$/5500 \AA)$^{-0.7}$.
The effect of dust on the colours is  
indicated by an orange arrow on 
the figures in section \ref{sect:models_results}. 
The length of the
arrow corresponds to a galaxy with                   
$\hat{\tau}_{V}$ = 2 and $\mu$ = 0.5, corresponding to 
an attenuation of 2.2 mag in the V-band for stars younger than 10$^{7}$ yr and 
of 1.1 mag for older stars. This is a reasonable upper limit to attenuation by dust. 
The cross on the arrow indicates the amount of 
reddening that is already included in the models.

\subsection{Difference between the BC03 and preliminary CB07 models}\label{sect:bc03_cb07}

The preliminary version of the CB07 model used here differs from the BC03 
models only in the treatment of AGB stars. The new models include the prescription
of \citet{Marigo2007} for the TP-AGB evolution of low- and intermediate-mass
stars. This prescription includes several important theoretical improvements 
over previous calculations, and it has been calibrated using carbon star
luminosity functions in the Magellanic Clouds and TP-AGB lifetimes (star
counts) in Magellanic Cloud clusters (we refer to the paper of \citet{Marigo2007}
for detail). As outlined by \citet{Bruzual2007}, this implies brighter $K$-band 
magnitudes and redder near-IR colours for the preliminary CB07 model than
for the BC03 model. 
We caution the reader that the spectra of TP-AGB stars and the stellar
evolution prescription for pre-AGB phases have not been updated in the 
preliminary version of CB07 model used here. The final version of this model
will include new spectra for TP-AGB stars and an updated library of 
\citet{Marigo2007} TP-AGB calculations connected to a new comprehensive grid 
of pre-AGB stellar evolutionary tracks by Bertelli et al. (2007, in preparation).
Nevertheless we feel that this preliminary analysis is informative. 

\begin{figure}
\begin{center}
\includegraphics[width=0.5\textwidth]{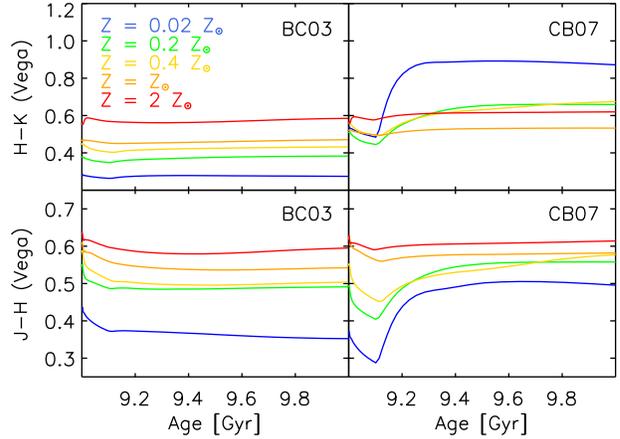}
\caption{Evolution of the near-IR colours predicted by the BC03 and CB07 stellar population models following 
a burst of star formation. 
Blue, green, yellow, orange and red lines
correspond to models with 0.02$Z_{\odot}$, 0.2$Z_{\odot}$, 
0.4$Z_{\odot}$, $Z_{\odot}$ and 2$Z_{\odot}$
respectively.}
\label{fig:fig3}
\end{center}
\end{figure}

To illustrate the differences between the BC03 and CB07 models,
we have computed the
evolution of the near-IR colours after a burst of star formation.
We assume that
one third of the final mass of the galaxy is formed during the burst, 
which occurs at $t$ = 9 Gyr on top of a galaxy with exponentially 
declining SFR.
As seen from Figure 
\ref{fig:fig3}, when TP-AGB stars appear, about 0.1 Gyr after the 
burst, the CB07 models are significantly redder (from a tenth to a few tenths
of a magnitude) in $J-H$ and $H-K$ than the BC03 model.  
This effect is seen mainly at metallicities of
0.2$Z_{\odot}$ and below. At such metallicities, the TP-AGB (and in particular, 
luminous carbon stars) dominate the near-IR light of intermediate-age stellar
populations in the CB07 model, while RGB stars (which are less luminous than
TP-AGB stars) prevail at medium to high metallicities.  

\subsection{Results from BC03 and preliminary CB07 models}\label{sect:models_results}

Figures \ref{fig:fig4}, \ref{fig:fig5} and 
\ref{fig:fig6}, \ref{fig:fig7} present
comparisons of the BC03 and preliminary CB07 models with our data. The
observational data are plotted as
black dots in four colour-colour diagrams: $g-r$ versus $u-g$, $J-H$ 
versus $Y-J$, $Y-K$ versus $H-K$ and $g-r$ versus $Y-K$. 
The model grids are defined by 3 different ages and 5 
metallicities from 0.02$Z_{\odot}$ to 2$Z_{\odot}$. 
The dotted, 
continuous and dashed-dotted lines correspond to model ages of 5 , 10 and 15 Gyr,
respectively. For the star-forming sample, the models with
constant SFR are plotted as stars, while the models with star formation
timescale $\tau$ = 3 Gyr are plotted as filled squares
(Figs.\ref{fig:fig4} and \ref{fig:fig5}). 
For the non star-forming sample, all models have $\tau$ = 1 Gyr and are
plotted as filled triangles (Figs.\ref{fig:fig6} and 
\ref{fig:fig7}).
All models in Figs.\ref{fig:fig4}--\ref{fig:fig7}
include dust attenuation with           
$\hat{\tau}_{V}$ = 1 and $\mu$ = 0.3. The orange arrow on the lower right corner
of each panel indicates the amount of reddening for a galaxy with        
$\hat{\tau}_{V}$ = 2 and $\mu$ = 0.5. The orange
cross indicates the reddening that is already included in our models. 

\subsubsection{Star-forming galaxies}

\begin{figure*}
\begin{center}
\includegraphics[width=0.75\textwidth]{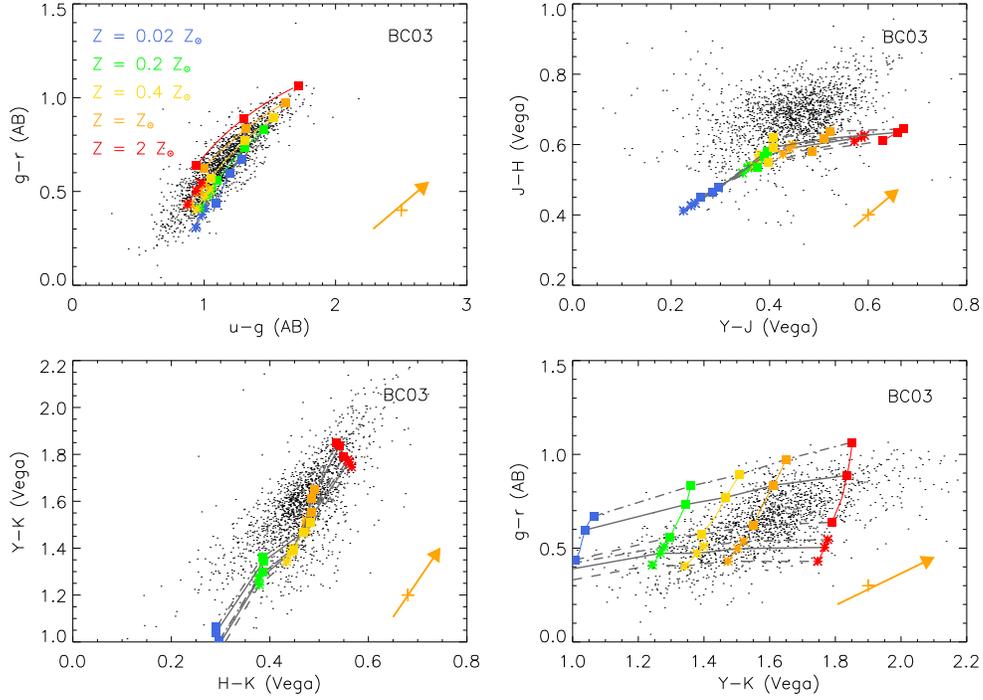}
\caption{Colour-colour diagrams showing BC03 model tracks
superimposed on our sample of star-forming galaxies (black dots). The stars and the
squares denote models with constant star-formation and exponentially declining star
formation with
$\tau$ = 3 Gyr. The grids are drawn for 5 metallcities and three
ages: the coloured lines represent isometallicity colours of 0.2$Z_{\odot}$, 
0.4$Z_{\odot}$,$Z_{\odot}$, 2$Z_{\odot}$ and the dotted, continuous and
dashed-dotted line show isochrones of 5, 10 and 15 Gyr respectively. The orange 
arrow indicates the amount of reddening for a galaxy with              
$\hat{\tau}_{V}$ = 2 and $\mu$ = 0.5. The orange
cross shows the reddening that is already included in our models.}
\label{fig:fig4}
\end{center}
\end{figure*}

\begin{figure*}
\begin{center}
\includegraphics[width=0.75\textwidth]{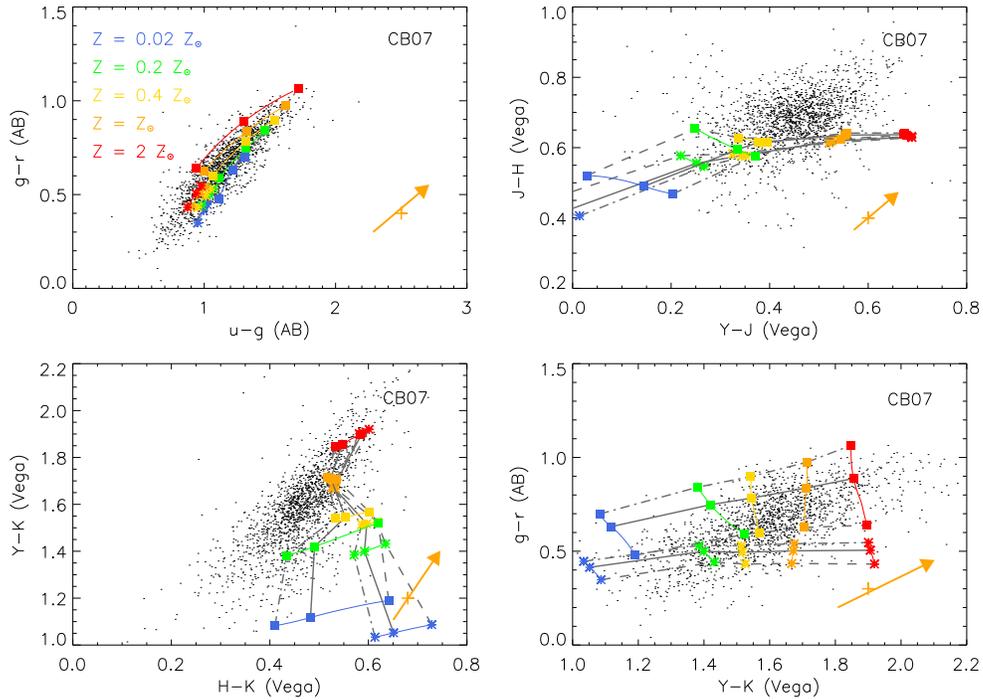}
\caption{As in Fig. \ref{fig:fig4}, except for the CB07 models.}              
\label{fig:fig5}
\end{center}
\end{figure*}

As seen from Fig. \ref{fig:fig4}, the BC03 models
cover the data in the optical $g-r$ versus $u-g$ colour-colour plane. 
The very bluest 
galaxies in $u-g$ can be fitted if a burst of recent star
formation is superposed on the star formation history used in our models. The 
colours of the reddest galaxies can be understood if they are significantly 
reddened by dust. 
We therefore conclude that our choices of star formation history, age and
metallicity are appropriate for the star-forming sample. These models should
then be able to reproduce the near-IR colours of the galaxies in our sample.

We find that, however, the BC03 models fail to cover the data 
in the $J-H$ versus $Y-J$  
and $Y-K$ versus $H-K$ colour-colour planes. 
In the near-IR, these models are sensitive primarily to metallicity.
As can be seen, the models cannot  
reproduce the observed scatter of the star-forming galaxies in the near-IR
colour-colour diagrams.                    
Neither reddening nor a scatter in age or star formation history
can account for the observed spread. We also note that the $J-H$ colours 
of the model tracks are too blue 
by approximatively 0.1 magnitude as compared to the data. 
This may not seem like a large 
offset, but as the near-IR 
colours of galaxies cover a small range of magnitude (0.4
magnitude in $J-H$) this is not a negligible discrepancy. 

Fig. \ref{fig:fig5} shows the comparison with the
preliminary version of the CB07 
models. As expected, the results 
do not differ from the BC03 models at optical wavelengths. On the
other hand, the near-IR colours are now dependent
on stellar age at metallicities
$<$ 0.4$Z_{\odot}$. Although this could in principle provide a mechanism for
understanding the observed scatter 
in the near-IR colour-colour planes, we find that
these more recent models still cannot fit the detailed trends seen in the data.   
The $H-K$ colours are skewed to significantly redder values in the CB07 models 
as compared to the data and the models fail to reproduce the
observed spread in $J-H$ colour at $Y-J$ colours greater than 0.4.
This conclusion is likely to change when new spectra are adopted for TP-AGB 
stars in the final models. The $Y-J$ and $Y-K$ colours are better reproduced
by the models. Both these colours are predicted to be good indicators
of metallicity. In particular, the $Y-K$ colour has almost
no dependence on age in both the BC03 and the CB07 models.  We note that
there is an offset in the UKIDSS $Y$-band magnitudes of $\sim$0.09 in the sense that
they are too faint (P.Hewett, private communication). Correcting for this
effect will result in a better match in $Y-K$ colour.   

It is also interesting to consider what these 
preliminary CB07 models predict in terms of {\it trends}
in the near-IR colours as a function of age and metallicity.   
The $Y-J$ colour exhibits similar behaviour to optical colours, in that 
it increases with age. The $J-H$ 
colour also increases slightly with age at metallicities above 0.4$Z_{\odot}$, but 
it {\it decreases} with increasing age at lower metallicities.
This effect is even stronger for the $H-K$ colour: a 0.2
$Z_{\odot}$ galaxy becomes 0.2
magnitudes bluer as the galaxy
ages from 5 Gyr to 15 Gyr. 
The lower right panel 
of Fig. \ref{fig:fig5} shows the 
best combination of optical and near-IR colours for 
deriving age and metallicity deduced from these
preliminary CB07 models. As seen on
this plot, the effect of increasing age is to move the galaxy along the
$g-r$ colour axis, without affecting $Y-K$ colours.
Conversely, increasing metallicity moves the galaxy along the
$Y-K$ axis, with very little effect on the $g-r$ colour.

\subsubsection{Non star-forming galaxies}\label{sect:res_bc_nsf}

\begin{figure*}
\begin{center}
\includegraphics[width=0.8\textwidth]{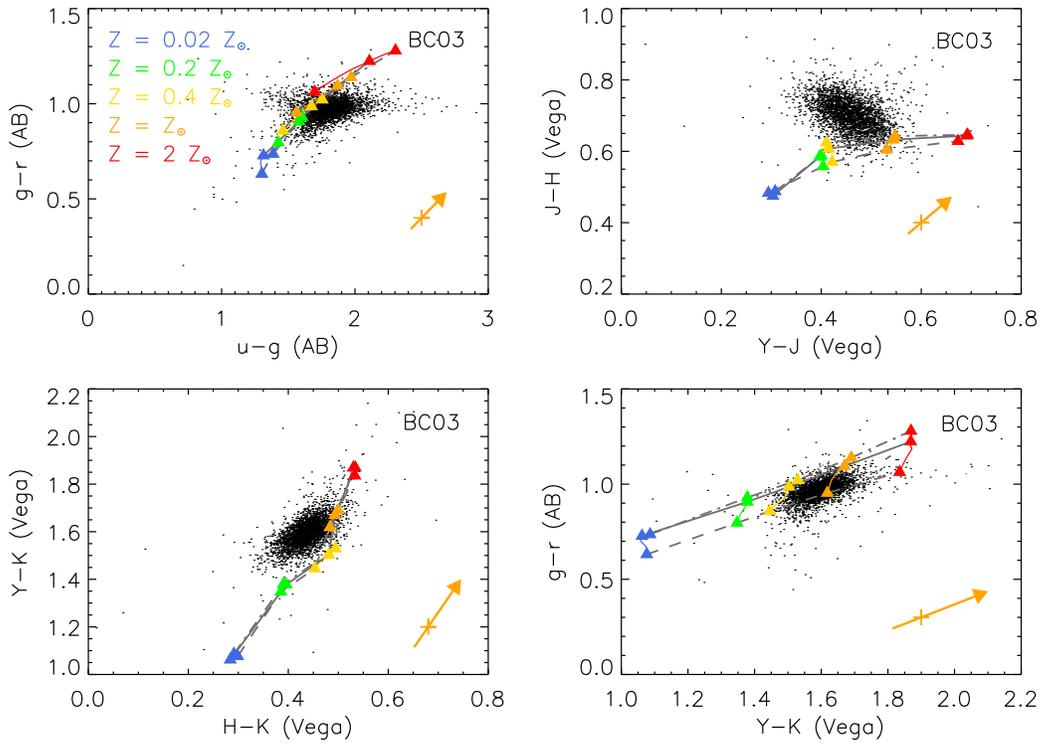}
\caption{As in Fig. \ref{fig:fig4}, 
except that the BC03 models are superimposed on our sample of non-star-forming
galaxies.}
\label{fig:fig6}
\end{center}
\end{figure*}

\begin{figure*}
\begin{center}
\includegraphics[width=0.8\textwidth]{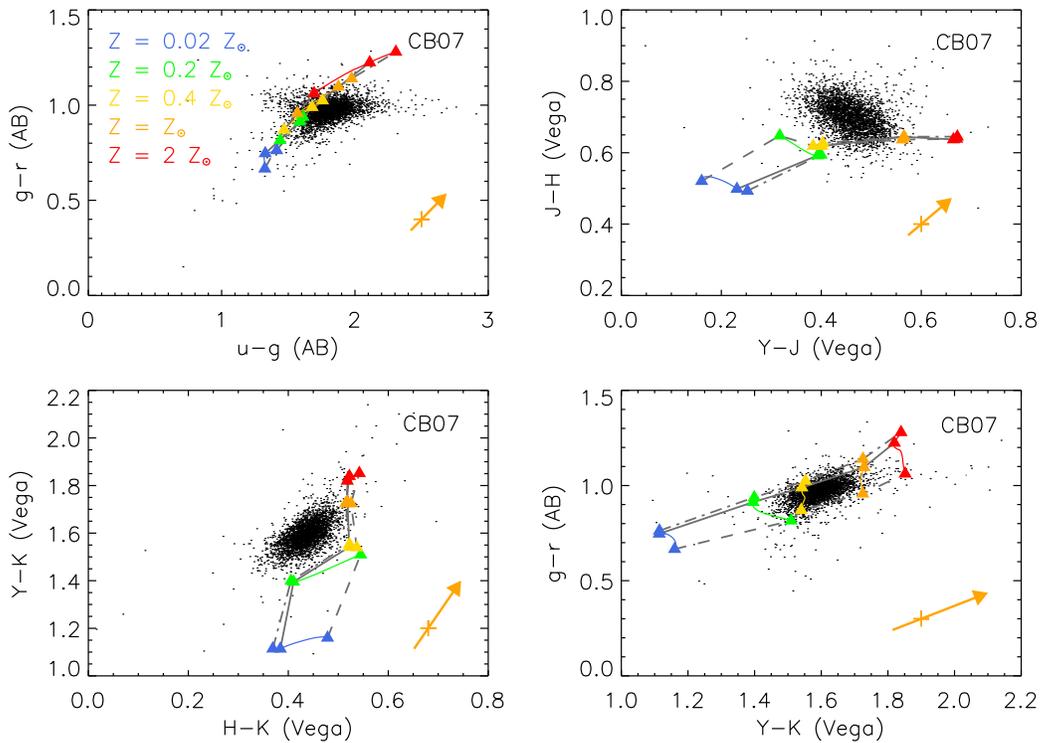}
\caption{As in Fig. \ref{fig:fig4}, except that the CB07 models are superimposed on our sample of 
non-star-forming galaxies.}
\label{fig:fig7}
\end{center}
\end{figure*}

In Fig. \ref{fig:fig6} we compare
the BC03 models with the colour distributions of 
galaxies in the non star-forming
sample. Recall that all the models on this plot have exponential declining
star formation histories with $\tau$ = 1 Gyr. The observed colours of 
non star-forming galaxies have significantly less
scatter than those of galaxies in the star-forming sample.
We find that models with metallicities between 0.4$Z_{\odot}$
and$Z_{\odot}$ reproduce the location of the observed galaxies in the
the $g-r$ versus $u-g$ colour-colour diagram, 
but the observed locus of the data in the $J-H$ versus $Y-J$ colour-colour
plane cannot be understood with these models.
The models also
predict $J-H$ colours that are too blue 
by about 0.1 magnitude. 

As seen from Fig. \ref{fig:fig7}, 
the preliminary CB07 models 
predict that the near-IR colours of non-star-forming galaxies 
depend weakly on age at metallicities below 0.4 
$Z_{\odot}$. The age dependence is less pronounced than for
the exponential declining
models with $\tau$ = 3 Gyr that we used to model star-forming galaxies.
This is due to the smaller fraction of luminous
TP-AGB stars relative to RGB stars in the
$\tau$ = 1 Gyr models at the ages considered here. 
As was the case for the star-forming galaxies,
we find that the preliminary version of the CB07 models do 
not match our data. 
These models are almost identical to the BC03
models at metallicities above 0.4$Z_{\odot}$. 
The galaxies in our non star-forming sample are mainly
massive, metal-rich systems, and the
low-metallicity models are not applicable to them.

\subsection{Comparison with \citet{Maraston2005} models} 

\subsubsection{Description of \citet{Maraston2005} models}\label{sect:des_M05}

One of the first papers to point out the influence of  
the short duration thermally pulsating (TP-)AGB phase on
the spectra of galaxies at $\lambda \sim 2 \mu m$ was written
by \citet{Maraston1998}.   
The most recent version of the Maraston models 
(\citealt{Maraston2005}; hereafter M05) implements this phase 
by adopting a `fuel
consumption' approach. In \citet{Rettura2006} M05 model 
predictions were compared with those of BC03 and PEGASE.2 
\citep{Fioc1997} for estimating the stellar mass content
of early-type galaxies at $z \sim 1$. 

In the following sections
we compare the predictions of CB07 with those of M05.   
In Figs. \ref{fig:fig8} and \ref{fig:fig9}, we plot
the M05 models and the preliminary 
CB07 models on the top of our star-forming sample and non 
star-forming sample in a series
of optical and near-IR colour-colour diagrams.
 For the star-forming sample, the star formation history is an exponentially
declining SFR with $\tau$ = 3 Gyr. 
The non star-forming sample is represented by an exponentially
declining SFR with $\tau$ = 1 Gyr.
Results are shown at three different
ages (5,10 and 15 Gyr), but only for half solar and solar
metallicities (we do not have lower metallicity
predictions for M05). Figure \ref{fig:fig9} shows
twice solar metallicity in addition.   
An extinction of $\tau_{V}$ = 0.3 is already 
included in the models. 
The arrow on each panel indicates the amount of extinction
corresponding to $\tau_{V}$ = 1.4 
as parametrised with a dust screen and the reddening curve of
\citet{Cardelli1989}, including
the update for the near-UV given by \citet{ODOnnell1994}.

\subsubsection{Results from M05 models for the star-forming
sample}\label{sect:results_M05_sf} 

\begin{figure*}
\centering
\includegraphics[width=6.6cm]{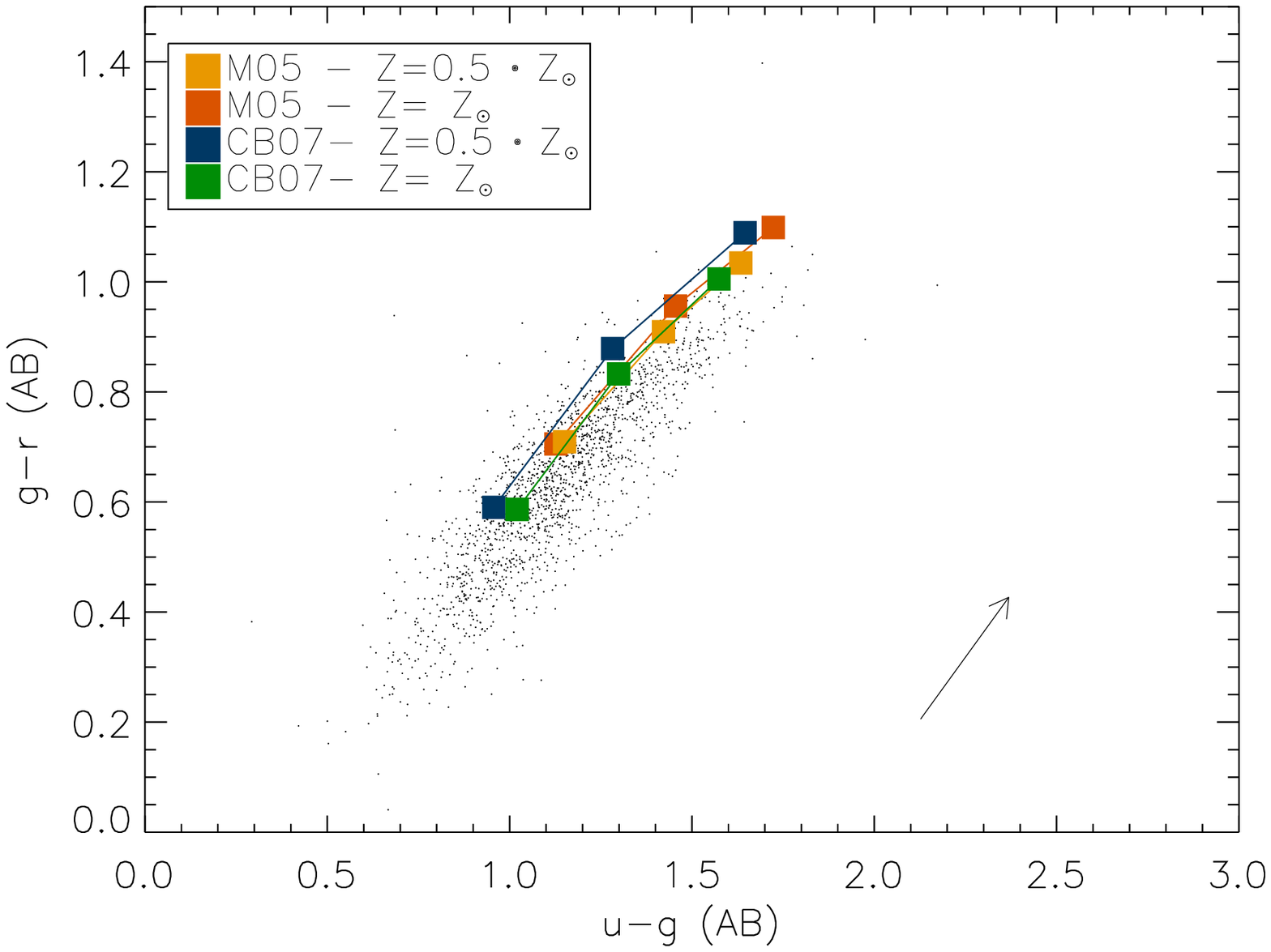}
\includegraphics[width=6.6cm]{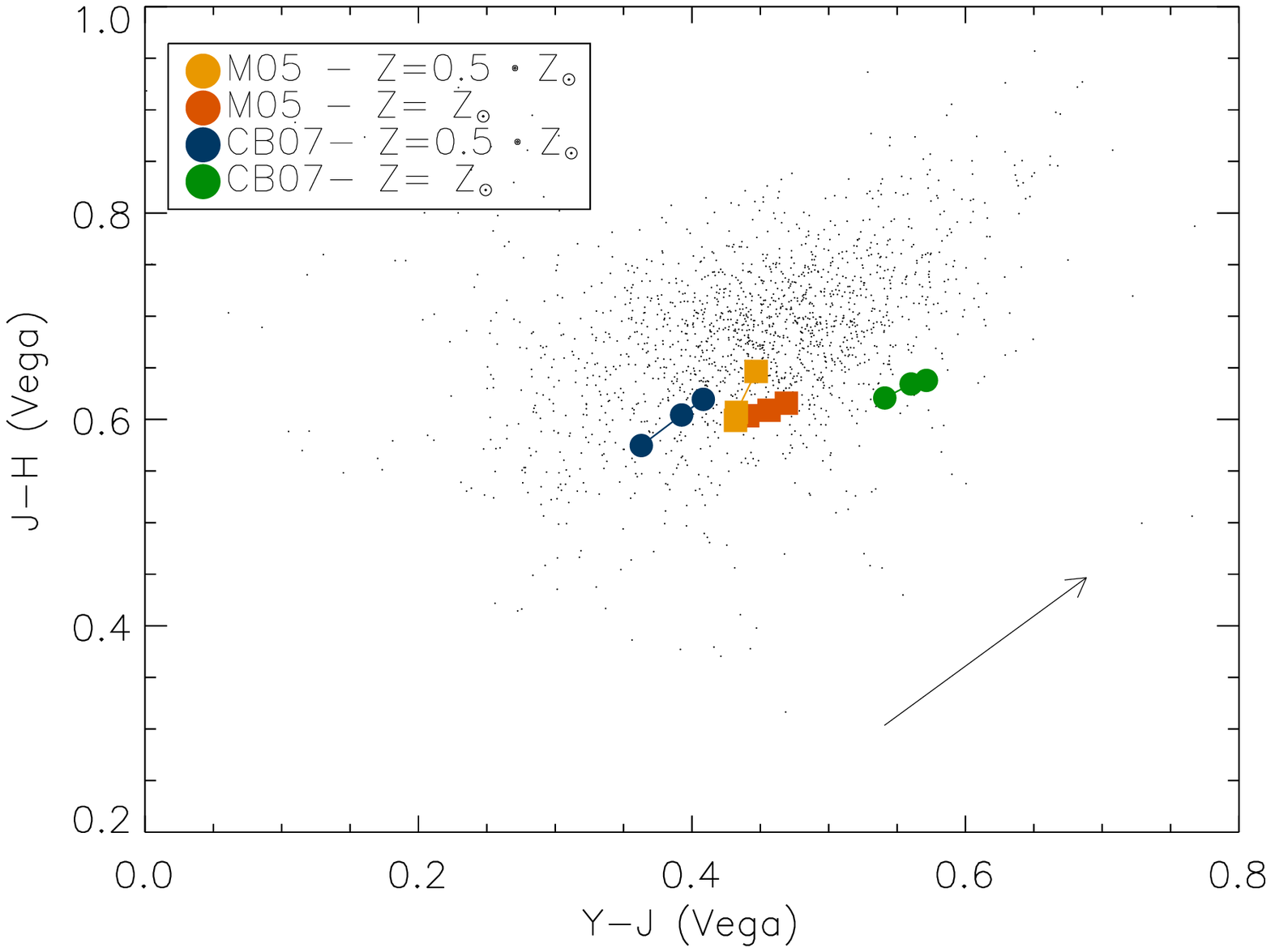}
\includegraphics[width=6.6cm]{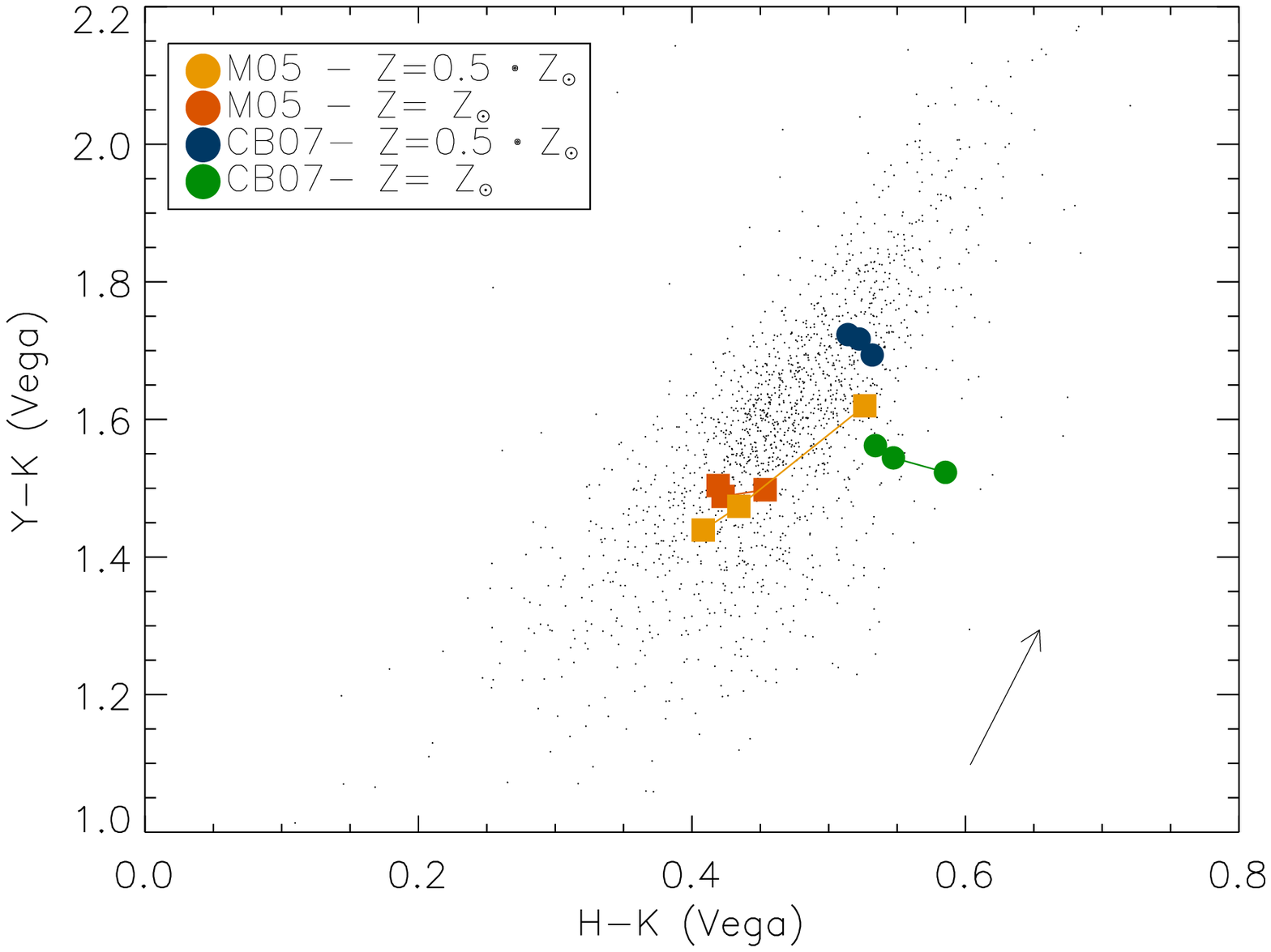}
\includegraphics[width=6.6cm]{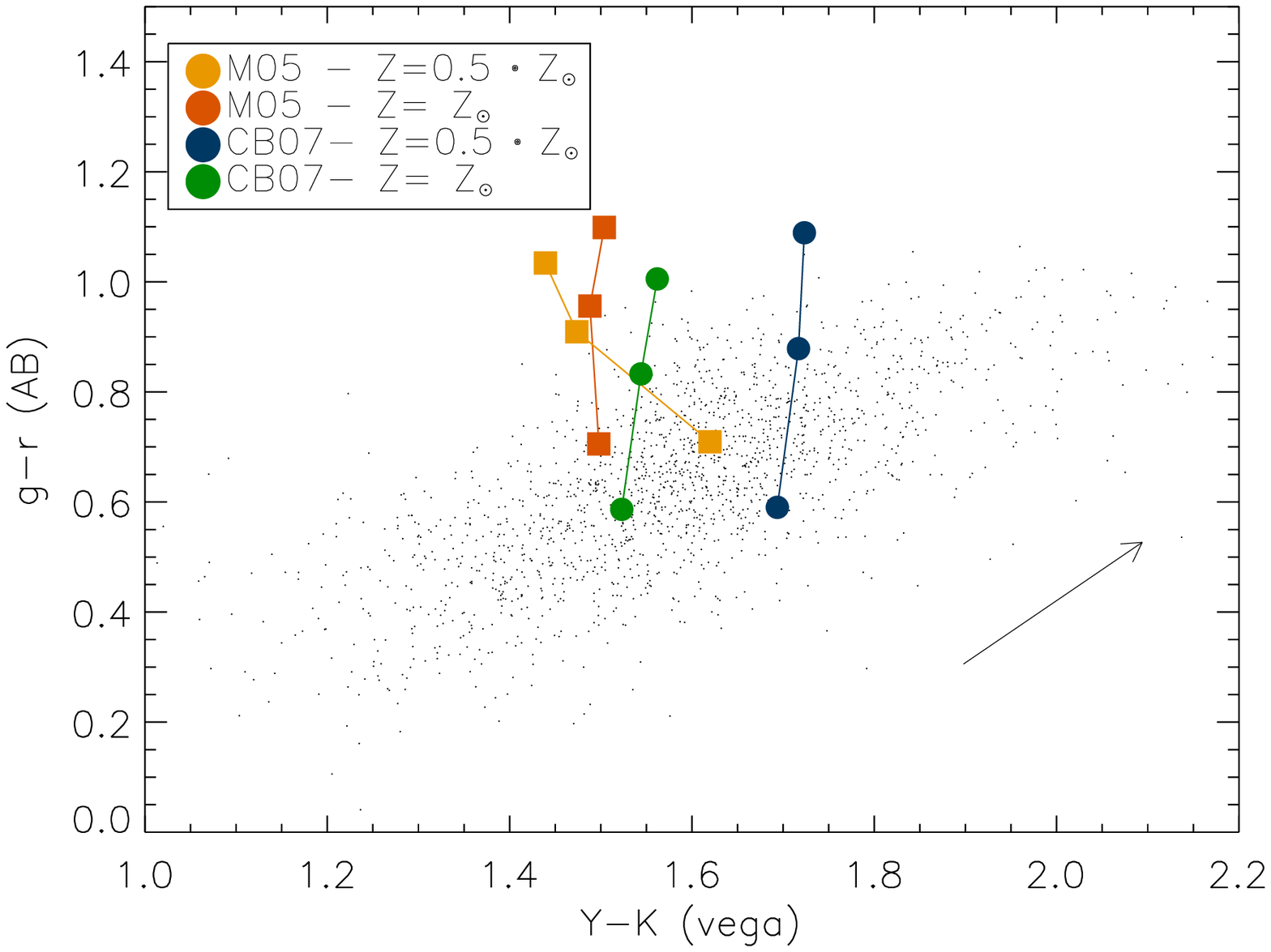}
\caption{Colour-colour plots of $M05$ and $CB07$ composite stellar population 
models superimposed on our star-forming galaxies in black dots. The squares 
account for the $M05$ exponential star-forming with $\tau = 3$ Gyr models. 
The circles account for the $CB07$ exponential star-forming with $\tau = 3$ 
Gyr models. The grids are drawn for $3$ metallicities and three ages 
($5, 10, 15$ Gyrs): the colourful lines represent isometallicity colours of 
0.5$Z_{\odot}$, $Z_{\odot}$. The amount of extinction labeled 
in the plots is parametrised with the reddening curve of \citet{Cardelli1989}, 
including the update for the near-UV given by \citet{ODOnnell1994}.}
\label{fig:fig8}
\end{figure*}

\begin{figure*}
\centering
\includegraphics[width=6.6cm]{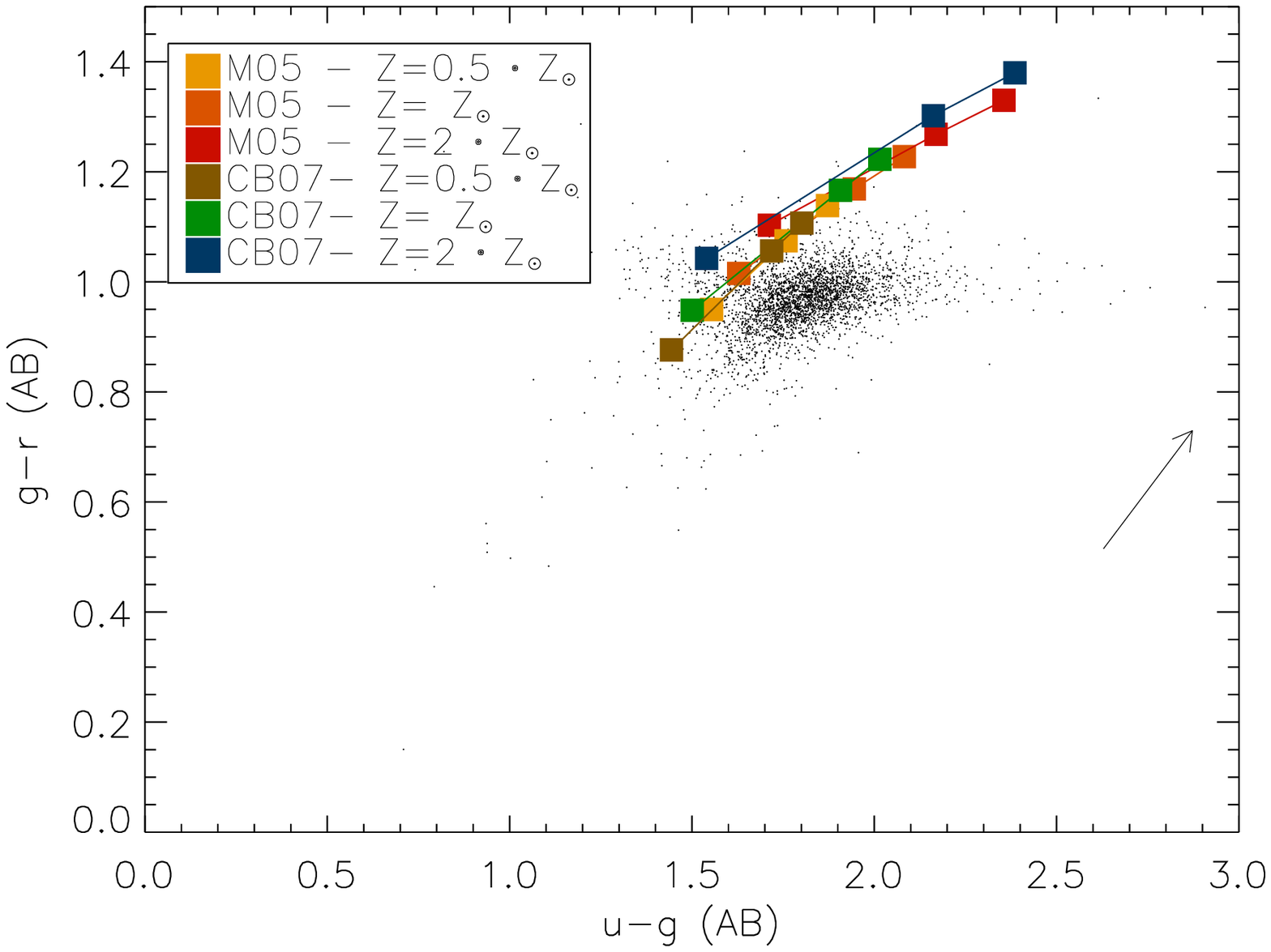}
\includegraphics[width=6.6cm]{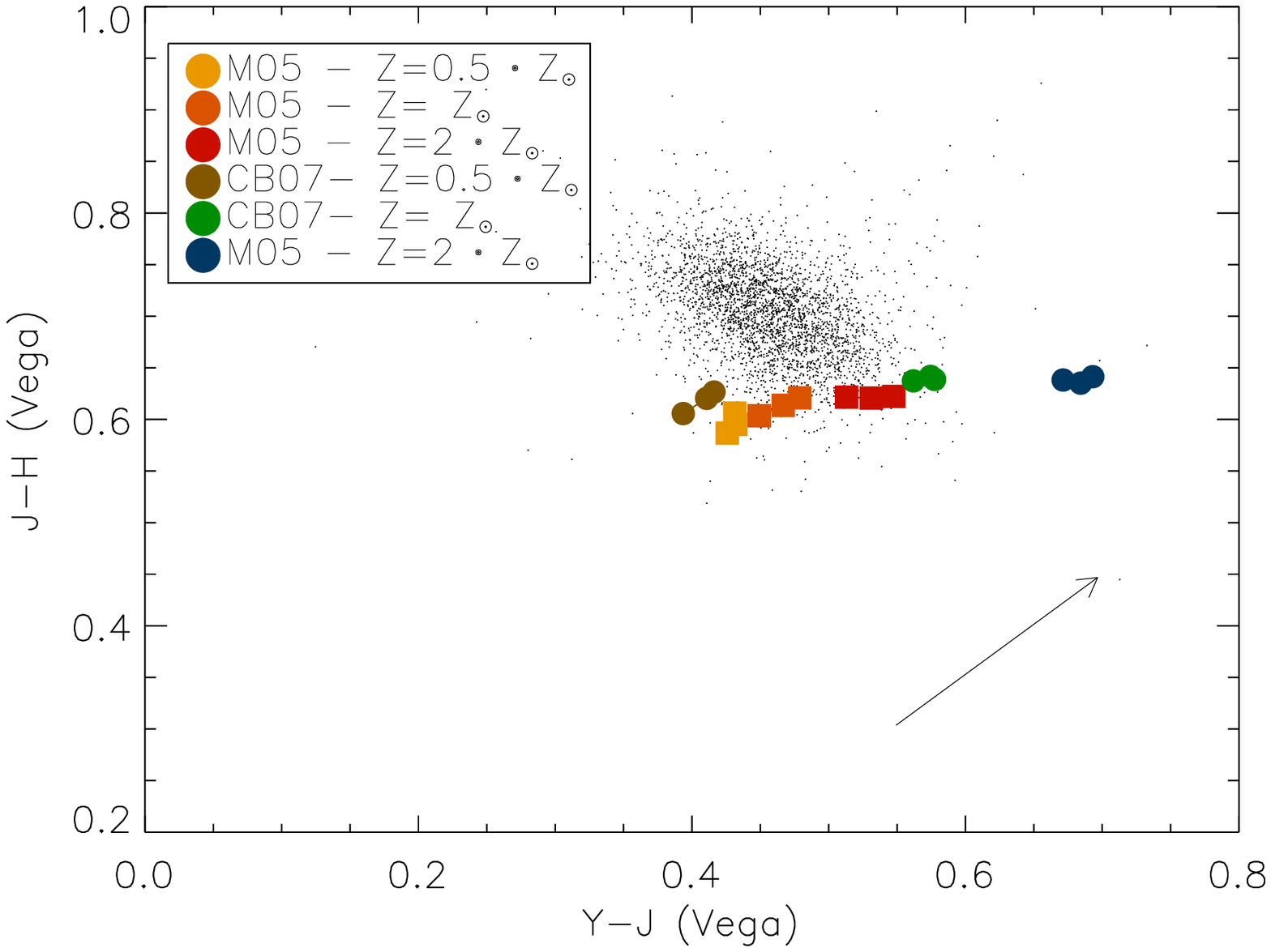}
\includegraphics[width=6.6cm]{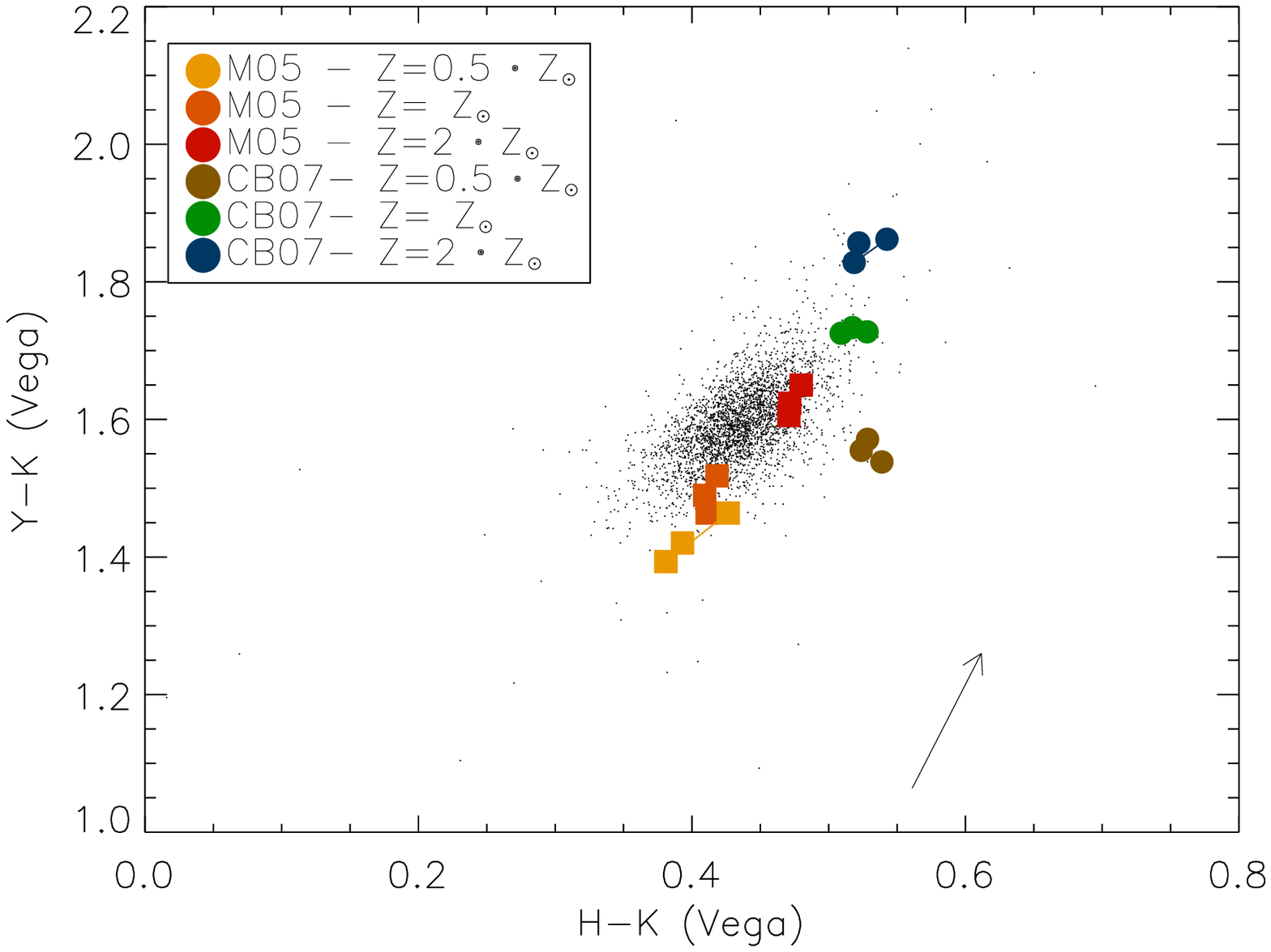}
\includegraphics[width=6.6cm]{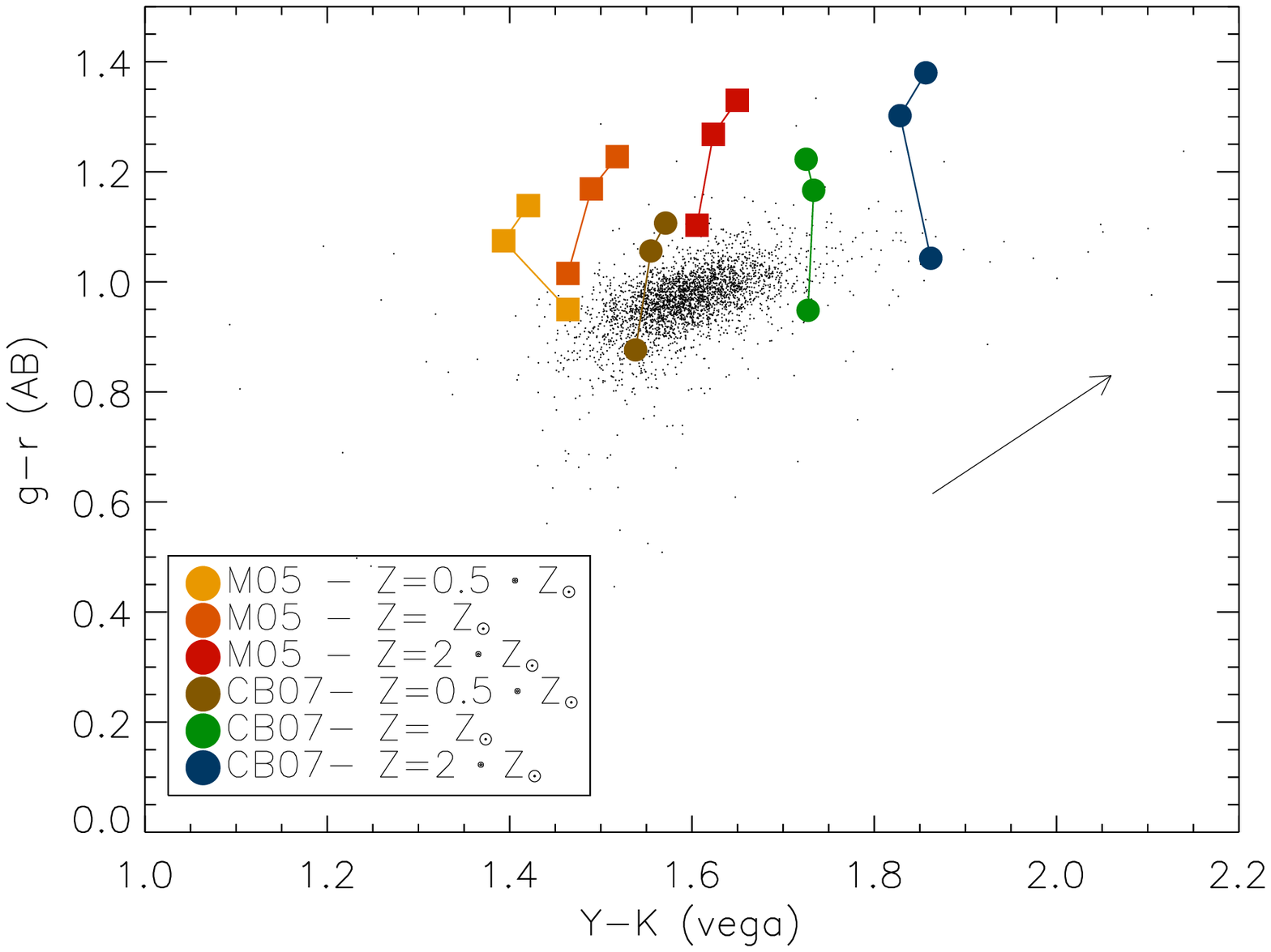}
\caption{As in Fig. \ref{fig:fig8}, except that the M05 and preliminary CB07 
models are superimposed on our sample of non-star-forming galaxies and that the
metallicities are 0.5$Z_{\odot}$, $Z_{\odot}$ and 2$Z_{\odot}$.}
\label{fig:fig9}
\end{figure*}

Figure \ref{fig:fig8} compares the predictions of CB07 with 
those of M05 for our star-forming sample, in the same way as described in the 
previous section. 
As seen from this figure, both models agree with the data for the optical
colours, M05 predicting slightly redder colours in $g-r$. In the $J-H$
vs $Y-J$ colour-colour plane, we notice that the preliminary CB07 models 
are considerably
more sensitive to metallicity. For both models, an extinction 
corresponding to $\tau_{V}$ = 1.4 is required to explain the reddest galaxies. 
This amount of extinction agrees with the measurement of the most dusty 
galaxies in our star-forming sample. On the other hand, such amount is not 
necessary to explain the distribution of the optical colours.  
The predictions of M05 models match the data particularly well in the 
$Y-K$ vs $H-K$ colour-colour plane; in
this plane, the preliminary CB07 models are offset in $H-K$. 
Both models exhibit a larger sensitivity to age at lower metallicity. 
The lower right panel of Fig. \ref{fig:fig8} shows that the M05 and CB07 
models differ
quite significantly in their predictions for $Y-K$ colours.
If one uses the preliminary CB07 model, one might infer that this colour is a
good indicator of
metallicity, but this is is not true for the M05 models. 
Based on this comparison, neither model is particularly favoured 
to explain the colour distribution of our star-forming sample.

\subsubsection{Results from M05 models for the non star-forming 
sample}\label{sect:results_M05_nsf} 

Figure \ref{fig:fig9} compares the predictions of CB07 with 
those of M05 for our non star-forming sample. 
 We have included the 2$Z_{\odot}$ isometallicity track as
these galaxies have measured stellar metallities in the range 0.7$Z_{\odot}$ 
to 1.5$Z_{\odot}$. We note that M05 models require systematically higher
metallicities (between $Z_{\odot}$ and 2$Z_{\odot}$) to fit
the observed colours than do the preliminary
CB07 models (between 0.5$Z_{\odot}$ and $Z_{\odot}$). The
dust reddening should barely affect the colours as the non star-forming galaxies
have a measured $z$-band attenuation close to zero.  
Our conclusions regarding the fit of the optical and near-IR
colours with both models are very similar to those reached for the star-forming sample. 
In optical colours, M05 models predict too red $g-r$ colours 
by approximately 0.2 magnitudes.
Both models predict too blue $J-H$ colour by around              
0.1 magnitude. The $H-K$ colour is better reproduced by 
M05 models.
As pointed out in section \ref{sect:res_bc_nsf}, 
the age dependence is less pronounced than for the exponential declining 
models that are used to model
the star-forming sample. The preliminary CB07 models provide a
better fit to the data in $Y-K$ colour; the 
M05 models requiring metallicities around 
2$Z_{\odot}$, which is larger than the metallicities
estimated for the galaxies in our sample \citep{Gallazzi2005}.


\section{Implications}\label{sect:discussion}

\subsection{Can the age-metallicity degeneracy be broken?}

As discussed in Section \ref{sect:models_results}, the $g-r$ versus $Y-K$
colour-colour plane may 
provide the ideal combination of colours
to break the age-metallicity degeneracy. 
How well does this work in practice?

\begin{figure}
\begin{center}
\includegraphics[width=0.48\textwidth]{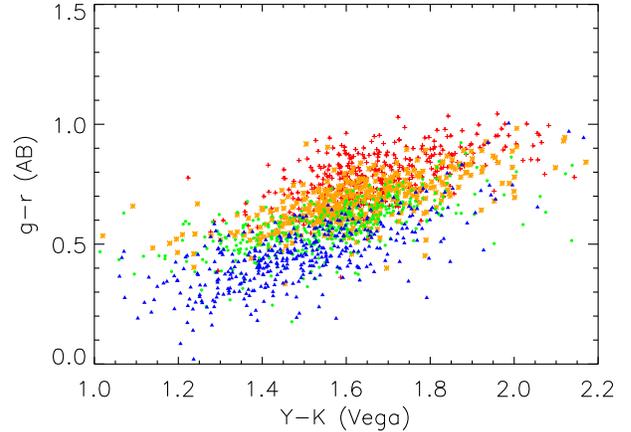}
\caption{Distribution of galaxies with different mean stellar ages in the 
$g-r$ versus $Y-K$ colour-colour plane.
The galaxies are divided in four equal classes
according to their stellar age. Blue (triangles), green (filled circles), 
orange (stars) and red (crosses) indicate
galaxies of increasing age.}
\label{fig:fig10}
\end{center}
\end{figure}

\begin{figure}
\begin{center}
\includegraphics[width=0.48\textwidth]{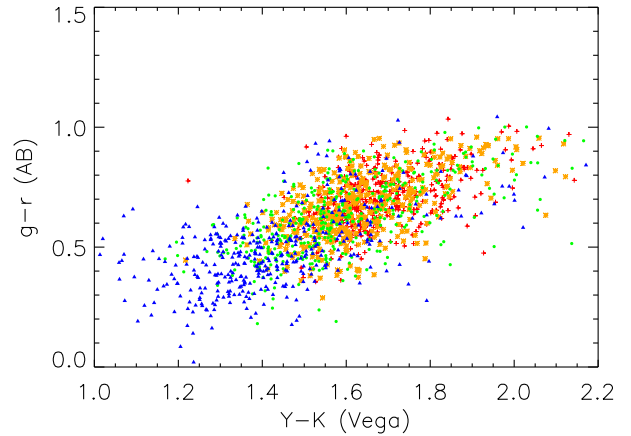}
\caption{As in Fig. 8, except galaxies are divided according to metallicity.}      
\label{fig:fig11}
\end{center}
\end{figure}

In this section we analyze how galaxies with different 
measurements of stellar age 
and gas-phase metallicity 
populate the $g-r$ versus $Y-K$ colour-color plane.  
Fig. \ref{fig:fig10} shows the distribution 
of galaxies with different mean stellar ages. We
have divided our sample into four age classes, which contain equal numbers of 
galaxies. Blue points represent the youngest 
galaxies and red points the oldest.
As seen from this Figure, the four age classes 
are distributed in horizontal slices, 
showing that as predicted by the models, the
$g-r$ colour is a good age indicator. 
Figure \ref{fig:fig11} shows how galaxies with different gas-phase
metallicities are distributed in the same colour-colour plane.
Again, we have divided our sample in four classes: the
blue crosses indicate the most metal-poor galaxies,
while the red crosses are for the most
metal rich ones. According to our preliminary version of
the CB07 models, the $Y-K$ colour should be very
sensitive to the metallicity. The data reveals that galaxies in the lowest metallicity
class do have predominantly blue $Y-K$ colours, but galaxies in the medium to high
metallicity classes all have similar $Y-K$ colours. 
Dust is probably the main
reason for this degeneracy. As seen from Fig. \ref{fig:fig5},
reddening can influence the $Y-K$ colours quite significantly, because the $Y$ and $K$
filters are distant in wavelength. 

In summary, our preliminary version of the 
CB07 models predict that the location of galaxies in the $g-r$
versus $Y-K$ colour-colour plane 
allows one to estimate age and metallicity independently for a fixed amount of
dust attenuation, but the data show that the method is not as clean as 
it might first appear.

\subsection{Estimating mass-to-light ratios}

\begin{figure}
\begin{center}
\includegraphics[width=0.48\textwidth]{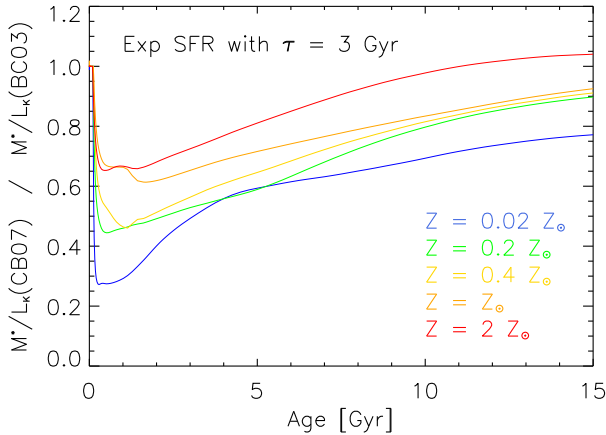}
\caption{K-band stellar mass-to-light ratio of CB07 model galaxies divided by the
K-band stellar mass-to-light
ratio of BC03 model galaxies is plotted as a function of age for an exponential declining 
SFR and for 5 metallicities from 0.02$Z_{\odot}$ to 2 
$Z_{\odot}$.}
\label{fig:fig12}
\end{center}
\end{figure}

When spectroscopic information is not available, the mass-to-light ratio of a galaxy can be
estimated by comparing the predicted spectral energy distributions of model 
galaxies derived using stellar population synthesis codes to the observational data.
The stellar mass of the galaxy is then obtained by multiplying its mass-to-light
ratio by its observed luminosity.
As discussed by \citet{deJong1996}, it is important that these mass estimates be
anchored by observations in at least one photometric bandpass 
that is not too sensitive to the presence of very young stars, or to 
metallicity and dust attenuation. The $K$-band is commonly regarded as the most reliable in
this respect, 
because dust and young stars should have little effect on the light emitted
by a galaxy at wavelengths around 2 $\mu$m.
Therefore, as pointed out by \citet{Maraston2006} and by
\citet{Bruzual2007}, the influence of AGB stars on the model predictions for the $K$-band luminosity
have very important implications for the reliability of stellar mass estimates 
that make use of data in this wavelength range. 
In Fig. \ref{fig:fig12}, we compare M$_{*}$/L$_{K}$ predicted by the BC03 and preliminary CB07
models as a function of the age of the galaxy. We show results for 5 different metallicities
for a model with exponential declining SFR with $\tau$ = 3
Gyr. As described in the previous section, this model provides a rather good fit to
the optical colours of the star-forming galaxies in our sample. 
As can be seen from this Figure, at ages less than 1 Gyr, the mass-to-light ratio 
predicted by the CB07 models 
ranges from 30\% (for 0.02$Z_{\odot}$) to 70\%
(for 2$Z_{\odot}$) of the value predicted by the BC03 models. The difference between the two models
decreases as a function of the age of the galaxy. At 10 Gyr, the typical age of the galaxies in our
sample, the mass predicted by CB07 model ranges from 70\% to 100\% of the value predicted by the BC03 model.
We conclude that even for present-day star-forming galaxies, there are significant
uncertainties on stellar masses derived from $K$-band luminosities that depend on
how AGB stars are treated in the population synthesis models.
The stellar masses predicted by the CB07 models are considerably smaller that those
predicted by the BC03 models, the effect being more important at high redshifts
and for low metallicity galaxies. 

\section{Summary}\label{sect:summary}

We have used a combination of empirical and theoretical techniques to 
interpret the near-IR colours of 
5800 galaxies drawn from the SDSS
main spectroscopic sample. This study focuses on the inner regions of galaxies
that are sampled by the 3 arcsecond SDSS fibre 
spectra. In the first part of our analysis, we study correlations between 
near-IR colours and physical parameters derived from the spectra, 
which include the specific SFR, the stellar age,
the metallicity, the dust attenuation and the axis ratio of the galaxy. 
All correlations are analyzed using galaxy samples that are matched in
redshift, stellar mass and concentration parameter. We analyze 
star-forming and non-star-forming galaxies separately.                
In the second part of our analysis, we compare the near-IR
colours of the galaxies in our sample to the predictions 
of stellar population models. Our main 
conclusions are as follows:

\begin {itemize}
\item Whereas more strongly star-forming galaxies have bluer optical colours,
the opposite is true at near-IR wavelengths: galaxies with higher values
of SFR/$M_*$ have redder near-IR colours.
This result agrees well with
the predictions of models in which TP-AGB stars dominate the $H$ and $K$-band light of
a galaxy following a burst of star formation. 

\item We find a surprisingly strong correlation between the near-IR colours and   
dust attenuation measured from the Balmer decrement.
However, the near-IR colours do not correlate with the axis ratios 
of the galaxies. This suggests that the correlation with dust attenuation
arises because TP-AGB stars are the main source of dust in the galaxy.  

\item We compare the near-IR colours of the galaxies in our sample 
with the colours predicted by the BC03 code, a preliminary version 
of the CB07 code and the M05 code. The preliminary CB07 code provides better 
{\it qualitative} agreement with
the data than BC03, in that it predicts that more strongly star-forming galaxies 
have redder near-IR colours. However, this effect occurs only at low metallicities.
The observed spread in the the near-IR colours of the star-forming 
galaxies is still unexplained by the current models.
The M05 models agree well with the data 
in the $Y-K$ vs $H-K$ 
colour-colour plane, but do not provide as a good a match for other   
colour combination.

\item The preliminary version of the CB07 model 
suggests that both the age and the metallicity of a galaxy
can be estimated from its location in the $g-r$ versus $Y-K$
colour-colour plane. We have tested this using our data and we find that
although age can be quite accurately estimated from the $g-r$
colour, the $Y-K$ colour can only be used to distinguish the lowest metallicity
galaxies.  

\item Even for present-day star-forming galaxies, the $K$-band mass-to-light ratios
predicted by the CB07 models can be significantly (up to 30\%) 
smaller than those predicted by the BC03 models.
This systematic uncertainty in the mass estimates is comparable to
the differences obtained when stellar masses are estimated
using a variety of different methods, for example 
fitting stellar population synthesis models to broadband colors  
or analysis of spectroscopic features \citep{Drory2004}.

Throughout this paper, we have been careful to stress that the model 
comparisons have been made using a preliminary version of the CB07 model.
The published version of the model should include an improved
library of spectra of TP-AGB stars, as well as new stellar evolutionary
tracks for the pre-AGB phase. This is likely to improve the agreement
with the data. The hope is that the sample of nearby galaxies with
high quality photometric and spectroscopic data from UKIDSS and SDSS
presented in this paper will continue to serve as a means of testing
and calibrating future stellar population synthesis models.

\end {itemize}


\section*{acknowledgements}
We thank Anna Gallazzi, Anthony Smith and Paula Coelho for useful discussions.
We are grateful to Paul Hewett for providing useful comments on UKIDSS data. 
CE aknowledges the Max Planck Institute for hospitality and support while this
work was carried out. CE was partly supported by the Swiss Sunburst Fund and the
Barbour fondation.

The United Kingdom Infrared Telescope is operated by the Joint Astronomy 
Centre on behalf of the UK Particle Physics and Astronomy Research Council. 
Funding for the SDSS has been provided by the Alfred P. Sloan Foundation, the 
Participating Institutions, the National Science Foundation, the US Department 
of Energy, the National Aeronautics and Space Administration, the Japanese 
Monbukagakusho, the Max Planck Society, and the Higher Education Funding 
Council for England. The SDSS Web site is http://www.sdss.org. The SDSS is 
managed by the Astrophysical Research Consortium for the Participating 
Institutions. The Participating Institutions are the American Museum of 
Natural History, the Astrophysical Institute Potsdam, the University of Basel, 
Cambridge University, Case Western Reserve University, the University of 
Chicago, Drexel University, Fermilab, the Institute for Advanced Study, the 
Japan Participation Group, Johns Hopkins University, the Joint Institute for 
Nuclear Astrophysics, the Kavli Institute for Particle Astrophysics and 
Cosmology, the Korean Scientist Group, the Chinese Academy of Sciences, Los 
Alamos National Laboratory, the Max Planck Institute for Astronomy, the Max 
Planck Institute for Astrophysics, New Mexico State University, Ohio State 
University, the University of Pittsburgh, the University of Portsmouth, 
Princeton University, the US Naval Observatory, and the University of 
Washington.


\bibliographystyle{mn2e} 
\bibliography{sam}

\begin{thebibliography}{}

\bibitem[\protect\citeauthoryear{{Aaronson}}{{Aaronson}}{1978}]{Aaronson1978}
{Aaronson} M.,  1978, ApJL, 221, L103

\bibitem[\protect\citeauthoryear{{Adelman-McCarthy et al.}}{{Adelman-McCarthy
  et al.}}{2006}]{Adelman_McCarthy2006}
{Adelman-McCarthy et al.} 2006, ApJS, 162, 38

\bibitem[\protect\citeauthoryear{{Baldwin}, {Phillips} \&
  {Terlevich}}{{Baldwin} et~al.}{1981}]{Baldwin1981}
{Baldwin} J.~A.,  {Phillips} M.~M.,    {Terlevich} R.,  1981, PASP, 93, 5

\bibitem[\protect\citeauthoryear{{Bell} \& {de Jong}}{{Bell} \& {de
  Jong}}{2000}]{Bell2000}
{Bell} E.~F.,  {de Jong} R.~S.,  2000, MNRAS, 312, 497

\bibitem[\protect\citeauthoryear{{Blanton et al.}}{{Blanton et
  al.}}{2003}]{Blanton2003}
{Blanton et al.} 2003, AJ, 125, 2348

\bibitem[\protect\citeauthoryear{{Brinchmann}, {Charlot}, {White}, {Tremonti},
  {Kauffmann}, {Heckman} \& {Brinkmann}}{{Brinchmann}
  et~al.}{2004}]{Brinchmann2004}
{Brinchmann} J.,  {Charlot} S.,  {White} S.~D.~M.,  {Tremonti} C.,  {Kauffmann}
  G.,  {Heckman} T.,    {Brinkmann} J.,  2004, MNRAS, 351, 1151

\bibitem[\protect\citeauthoryear{{Bruzual}}{{Bruzual}}{2007}]{Bruzual2007}
{Bruzual} G.,  2007, (astro-ph/0703052)

\bibitem[\protect\citeauthoryear{{Bruzual} \& {Charlot}}{{Bruzual} \&
  {Charlot}}{2003}]{Bruzual2003}
{Bruzual} G.,  {Charlot} S.,  2003, MNRAS, 344, 1000

\bibitem[\protect\citeauthoryear{{Cardelli}, {Clayton} \& {Mathis}}{{Cardelli}
  et~al.}{1989}]{Cardelli1989}
{Cardelli} J.~A.,  {Clayton} G.~C.,    {Mathis} J.~S.,  1989, ApJ, 345, 245

\bibitem[\protect\citeauthoryear{{Charlot} \& {Bruzual}}{{Charlot} \&
  {Bruzual}}{2007}]{Charlot2007}
{Charlot} S.,  {Bruzual} G.,  2007, in preparation

\bibitem[\protect\citeauthoryear{{Charlot} \& {Fall}}{{Charlot} \&
  {Fall}}{2000}]{Charlot2000}
{Charlot} S.,  {Fall} S.~M.,  2000, ApJ, 539, 718

\bibitem[\protect\citeauthoryear{{De Jong}}{{De Jong}}{1996}]{deJong1996}
{De Jong} R.~S.,  1996, A\&AS, 118, 557

\bibitem[\protect\citeauthoryear{{Drory}, {Bender} \& {Hopp}}{{Drory}
  et~al.}{2004}]{Drory2004}
{Drory} N.,  {Bender} R.,    {Hopp} U.,  2004, ApJL, 616, L103

\bibitem[\protect\citeauthoryear{{Fioc} \& {Rocca-Volmerange}}{{Fioc} \&
  {Rocca-Volmerange}}{1997}]{Fioc1997}
{Fioc} M.,  {Rocca-Volmerange} B.,  1997, A\&A, 326, 950

\bibitem[\protect\citeauthoryear{{Frogel}}{{Frogel}}{1985}]{Frogel1985}
{Frogel} J.~A.,  1985, ApJ, 298, 528

\bibitem[\protect\citeauthoryear{{Gallazzi}, {Charlot}, {Brinchmann}, {White}
  \& {Tremonti}}{{Gallazzi} et~al.}{2005}]{Gallazzi2005}
{Gallazzi} A.,  {Charlot} S.,  {Brinchmann} J.,  {White} S.~D.~M.,
  {Tremonti} C.~A.,  2005, MNRAS, 362, 41

\bibitem[\protect\citeauthoryear{{Geller}, {Kenyon}, {Barton}, {Jarrett} \&
  {Kewley}}{{Geller} et~al.}{2006}]{Geller2006}
{Geller} M.~J.,  {Kenyon} S.~J.,  {Barton} E.~J.,  {Jarrett} T.~H.,    {Kewley}
  L.~J.,  2006, AJ, 132, 2243

\bibitem[\protect\citeauthoryear{{Kauffmann et al.}}{{Kauffmann et
  al.}}{2003a}]{Kauffmann2003}
{Kauffmann et al.} 2003a, MNRAS, 341, 33

\bibitem[\protect\citeauthoryear{{Kauffmann et al.}}{{Kauffmann et
  al.}}{2003b}]{Kauffmann2003b}
{Kauffmann et al.} 2003b, MNRAS, 341, 54

\bibitem[\protect\citeauthoryear{{Kennicutt}
  Jr.}{{Kennicutt}}{1998}]{Kennicutt1998}
{Kennicutt} Jr. R.~C.,  1998, ARA\&A, 36, 189

\bibitem[\protect\citeauthoryear{{Lawrence et al.}}{{Lawrence et
  al.}}{2007}]{Lawrence2007}
{Lawrence et al.} 2007, MNRAS, 379, 1599

\bibitem[\protect\citeauthoryear{{Lee}, {Worthey}, {Trager} \& {Faber}}{{Lee}
  et~al.}{2007}]{Lee2007}
{Lee} H.-c.,  {Worthey} G.,  {Trager} S.~C.,    {Faber} S.~M.,  2007, ApJ, 664,
  215

\bibitem[\protect\citeauthoryear{{MacArthur}, {Courteau}, {Bell} \&
  {Holtzman}}{{MacArthur} et~al.}{2004}]{MacArthur2004}
{MacArthur} L.~A.,  {Courteau} S.,  {Bell} E.,    {Holtzman} J.~A.,  2004,
  ApJS, 152, 175

\bibitem[\protect\citeauthoryear{{Maraston}}{{Maraston}}{1998}]{Maraston1998}
{Maraston} C.,  1998, MNRAS, 300, 872

\bibitem[\protect\citeauthoryear{{Maraston}}{{Maraston}}{2005}]{Maraston2005}
{Maraston} C.,  2005, MNRAS, 362, 799

\bibitem[\protect\citeauthoryear{{Maraston}, {Daddi}, {Renzini}, {Cimatti},
  {Dickinson}, {Papovich}, {Pasquali} \& {Pirzkal}}{{Maraston}
  et~al.}{2006}]{Maraston2006}
{Maraston} C.,  {Daddi} E.,  {Renzini} A.,  {Cimatti} A.,  {Dickinson} M.,
  {Papovich} C.,  {Pasquali} A.,    {Pirzkal} N.,  2006, ApJ, 652, 85

\bibitem[\protect\citeauthoryear{{Marigo} \& {Girardi}}{{Marigo} \&
  {Girardi}}{2007}]{Marigo2007}
{Marigo} P.,  {Girardi} L.,  2007, A\&A, 469, 239

\bibitem[\protect\citeauthoryear{{Mobasher}, {Ellis} \& {Sharples}}{{Mobasher}
  et~al.}{1986}]{Mobasher1986}
{Mobasher} B.,  {Ellis} R.~S.,    {Sharples} R.~M.,  1986, MNRAS, 223, 11

\bibitem[\protect\citeauthoryear{{O'Donnell}}{{O'Donnell}}{1994}]{ODOnnell1994}
{O'Donnell} J.~E.,  1994, ApJ, 422, 158

\bibitem[\protect\citeauthoryear{{Rettura et al.}}{{Rettura et
  al.}}{2006}]{Rettura2006}
{Rettura et al.} 2006, A\&A, 458, 717

\bibitem[\protect\citeauthoryear{{Schlegel}, {Finkbeiner} \&
  {Davis}}{{Schlegel} et~al.}{1998}]{Schlegel1998}
{Schlegel} D.~J.,  {Finkbeiner} D.~P.,    {Davis} M.,  1998, ApJ, 500, 525

\bibitem[\protect\citeauthoryear{{Skrutskie et al.}}{{Skrutskie et
  al.}}{2006}]{Skrutskie2006}
{Skrutskie et al.} 2006, AJ, 131, 1163

\bibitem[\protect\citeauthoryear{{Tremonti et al.}}{{Tremonti et
  al.}}{2004}]{Tremonti2004}
{Tremonti et al.} 2004, ApJ, 613, 898

\bibitem[\protect\citeauthoryear{{Warren et al.}}{{Warren et
  al.}}{2007}]{Warren2007}
{Warren et al.} 2007, MNRAS, 375, 213

\bibitem[\protect\citeauthoryear{{Wu}, {Shao}, {Mo}, {Xia} \& {Deng}}{{Wu}
  et~al.}{2005}]{Wu2005}
{Wu} H.,  {Shao} Z.,  {Mo} H.~J.,  {Xia} X.,    {Deng} Z.,  2005, ApJ, 622, 244

\bibitem[\protect\citeauthoryear{{York et al.}}{{York et al.}}{2000}]{York2000}
{York et al.} 2000, AJ, 120, 1579

\end{thebibliography}

\end{document}